%% file: main.tex
\documentclass[11pt]{article}
\usepackage[margin=1in]{geometry}
\usepackage[utf8]{inputenc}
\usepackage[english]{babel}
\usepackage[dvipsnames]{xcolor}
\usepackage[normalem]{ulem}
\usepackage[labelfont=bf]{caption}
\usepackage{microtype, natbib, fancyhdr, siunitx, graphicx, hyperref, epstopdf, tikz, adjustbox, mathtools, float, anyfontsize}
\usepackage{bm, amssymb, amsmath, amsthm}
\usepackage{etoolbox}
\input{preamble.tex} 

\newbool{anon}
\setbool{anon}{false}

\newbool{figs}
\setbool{figs}{true}

\def\spacingset#1{\renewcommand{\baselinestretch}%
{#1}\small\normalsize} \spacingset{1}

\title{
  \textbf{Fast Adaptive Fourier Integration for Spectral Densities of Gaussian Processes} 
} \ifbool{anon}{}{ \author{ Paul G. Beckman\thanks{corresponding author:
\texttt{paul.beckman@cims.nyu.edu}} \thanks{PGB is supported in part by the
Office of Naval Research under award \#N00014-21-1-2383 and by the U.S.
Department of Energy, Office of Science, Office of Advanced Scientific Computing
Research, Department of Energy Computational Science Graduate Fellowship under
Award Number DE-SC0022158} \\ 
    \small Courant Institute, \\ 
    \small New York University \\ 
    \small New York, NY 10012 \\
    \and 
    Christopher J. Geoga \\
    \small Department of Statistics, \\ 
    \small University of Wisconsin-Madison \\
    \small Madison, WI 53706
    } }
\date{}

\begin{document}

\maketitle

\begin{abstract}
  \input{./src/abstract.tex}

  \bigskip

\end{abstract}

\newpage

\section{Introduction} \label{sec:intro}
\input{./src/intro.tex}

\section{Methodology} \label{sec:method}
\input{./src/method.tex}

\section{Numerical Results} \label{sec:results}
\input{./src/results.tex}

\section{Application to Doppler LiDAR data} \label{sec:lidar}
\input{./src/lidar.tex}
\section{Discussion} \label{sec:discussion}
\input{./src/discussion.tex}

\ifbool{anon}{}{
  \renewcommand{\abstractname}{Acknowledgments}
  \begin{abstract}
  \noindent The authors would like to thank Michael L. Stein and Michael O'Neil
  for their helpful suggestions and feedback.
  \end{abstract}

  \renewcommand{\abstractname}{Competing Interests}
  \begin{abstract}
  \noindent The authors report no competing interests.
  \end{abstract}
}

\bibliography{refs_firstinitial}{}

\newpage 

\appendix \section*{Appendix: Proofs} \label{sec:appendix}
\input{./src/proofs_supp.tex}

\end{document}


\maketitle

\renewcommand{\thesubsection}{S.\arabic{subsection}}

\input{./src/numerics_supp.tex}

\newpage

\input{./src/lidar_supp.tex}

%% file: preamble.tex
\newcommand{\E}{\mathbb{E}}    
\newcommand{\Z}{\mathbb{Z}}    
\newcommand{\R}{\mathbb{R}}    
\newcommand{\bO}{\mathcal{O}}  
\newcommand{\Fr}{\mathcal{F}}  
\newcommand{\lam}{\lambda}     
\newcommand{\omg}{\omega}      
\newcommand{\by}{\bm{y}}       
\newcommand{\bw}{\bm{\omega}}  
\newcommand{\bth}{\bm{\theta}} 
\newcommand{\bS}{\bm{\Sigma}}  
\newcommand{\bA}{\bm{A}}       
\newcommand{\abs}[1]{\left|#1\right|}      
\newcommand{\dif}{\mathop{}\!\mathrm{d}}   
\newcommand{\norm}[2][]{\left\Vert#2\right\Vert_{#1}} 
\newcommand{\shortnorm}[2][]{\Vert#2\Vert_{#1}}       
\newcommand{\mat}[1]{\begin{bmatrix}#1\end{bmatrix}}  

\renewcommand{\epsilon}{\varepsilon}

\newcommand{\expp}[1]{\exp(#1)}

\DeclareMathOperator*{\Cov}{Cov} 
\let\Re\relax \DeclareMathOperator*{\Re}{Re} 
 
\newcommand{\pder}[2]{\frac{\partial #1}{\partial #2}} 
\newcommand{\iid}{\mathrel{\overset{\scalebox{0.5}{\text{i.i.d.}}}{\scalebox{1.1}[1]{$\sim$}}}}

\newtheorem{theorem}{Theorem}
\newtheorem*{theorem*}{Theorem}
\newtheorem{lemma}{Lemma}
\newtheorem*{lemma*}{Lemma}

\newtheorem*{corollary*}{Corollary}

\newtheorem*{definition*}{Definition}

\newtheorem*{proposition*}{Proposition}
\theoremstyle{remark}
\newtheorem{remark}{Remark}
\newtheorem*{remark*}{Remark}

\numberwithin{equation}{section}

%% file: src/abstract.tex
The specification of a covariance function is of paramount importance when
employing Gaussian process models, but the requirement of positive definiteness
severely limits those used in practice. Designing flexible stationary covariance
functions is, however, straightforward in the spectral domain, where one needs
only to supply a positive and symmetric spectral density. In this work, we
introduce an adaptive integration framework for efficiently and accurately
evaluating covariance functions and their derivatives at irregular locations
directly from \textit{any} continuous, integrable spectral density. In order to
make this approach computationally tractable, we employ high-order panel
quadrature, the nonuniform fast Fourier transform, and a Nyquist-informed panel
selection heuristic, and derive novel algebraic truncation error bounds which
are used to monitor convergence. As a result, we demonstrate several orders of
magnitude speedup compared to naive uniform quadrature approaches, allowing us
to evaluate covariance functions from slowly decaying, singular spectral
densities at millions of locations to a user-specified tolerance in seconds on a
laptop. We then apply our methodology to perform gradient-based maximum
likelihood estimation using a previously numerically infeasible long-memory
spectral model for wind velocities below the atmospheric boundary layer.

%% file: src/intro.tex
Gaussian process (GP) models are ubiquitous in many statistical settings. They
provide a flexible method of interpolating data which incorporates dependence
structure, and yield model-implied second-order information that can be used for
uncertainty quantification. Let $Z(x)$ be a GP with mean function $\E Z(x) \equiv
0$ and positive definite covariance function
\begin{equation}
    \Cov\big(Z(x), Z(x')\big) = K_{\bth}(x, x'),
\end{equation}
where $K_{\bth}$ is some parametric family of positive definite covariance
functions indexed by parameters $\bth$. Consider observing $Z$ at locations
$x_1,...,x_n \in \R$ corresponding to measurements $y_i := Z(x_i)$. Then the
vector of observations $\by := [y_1,...,y_n] \in \R^n$ is distributed as
\begin{align}
    \by &\sim N\big(\bm{0}, \bS_{\bth}\big)
\end{align}
where the covariance matrix is given entry-wise by $(\bS_{\bth})_{ij} :=
K_{\bth}(x_i, x_j)$. For the duration of this work, we will make the additional
simplifying assumption that the process $Z$ is \emph{stationary}, so that
$K_{\bth}(x, x') = K_{\bth}(x - x')$, which implies that the process $Z(x)$ is
translation-invariant. This assumption limits the types of dependence structure
one can express in exchange for significant theoretical and computational
benefits. We refer readers to~\cite{stein1999interpolation} for a more detailed
introduction and discussion of GPs, covariance functions, and stationarity.

Once a parametric family of covariance models has been specified for some
dataset $\by$, practitioners often need to fit the model parameters $\bth$,
after which they can perform interpolation and other downstream tasks using the
law specified by the fitted parameters. In order to estimate the parameters
$\bth$, one can compute the maximum likelihood estimator (MLE), denoted
$\hat{\bth}$, which minimizes the Gaussian negative log-likelihood
\begin{equation}
    -2\ell(\bth) := \log\abs{\bS_{\bth}} + \by^\top \bS_{\bth}^{-1} \by + n\log(2\pi).
\end{equation}
Naturally, evaluating the log-likelihood in this form requires evaluation of
$K_{\bth}$ in order to construct $\bS_{\bth}$. If $K_{\bth}$ is available in
closed form, this is of course no issue. Common examples of covariance functions
in this category are the Mat\'ern model, with $K_{\bth}(r) \propto r^{\nu}
\mathcal{K}_{\nu}(r)$ and $\mathcal{K}_{\nu}$ the modified second-kind Bessel
function \citep{olver2010nist}, and its special cases $K_{\bth}(r) \propto
\expp{-|r|}$ for $\nu = 1/2$ and $K_{\bth}(r) \propto \expp{-|r|^2}$ for $\nu
\to \infty$. For the remainder of this paper we will often suppress the
dependence of $\bS, K,$ and $S$ on $\bth$ for notational clarity.

Unfortunately, as has been noted many times in the literature, positive definite
covariance functions that can be expressed in closed form are fairly difficult
to construct mathematically, and thus few are known and used in practice. This
severely limits the expressiveness of Gaussian process models available to
practitioners. For stationary processes, however, there is a much more flexible
way to construct valid covariance models by instead specifying their Fourier
transform. Bochner's theorem states that
\begin{equation} \label{eq:fourier-integral}
    K(r) 
    := \Fr\{S\}(r)
    := \int_{-\infty}^\infty S(\omega) \expp{2\pi i\omega r} \, d\omega
\end{equation}
is a positive definite function for \textit{any} integrable positive function
$S$, which we refer to as a \textit{spectral density}. Positive spectral
densities are much more easily and flexibly constructed and parameterized than
positive definite covariance functions. Thus from a modeling perspective it is
desirable to build and fit models via their spectral density.

Many methods for working in the spectral domain in a variety of special cases
already exist in the literature. For gridded data in one or two dimensions,
Whittle-type approximations are popular and fast to compute, but may introduce
severe bias without special care \citep{whittle1963stochastic,
sykulski2019debiased}. For spatio-temporal processes that are regular in at
least one dimension, ``half-spectral" models have been used to generate flexible
kernels
\citep{cressie1999classes,stein2005statistical,horrell2017half,geoga2021flexible}.
For gridded data in two dimensions with missing values,
\cite{guinness2019spectral} provides an imputation method for estimating
spectral densities via the EM algorithm. More generally applicable is the
``Random Fourier Features'' paradigm~\citep{rahimi2007random},which can be
viewed as a Monte Carlo approximation to the Fourier integral
(\ref{eq:fourier-integral}). A number of methods have also been proposed which
are equivalent to truncating the Fourier integral and applying the trapezoidal
rule on a finite interval, including the ``Regular Fourier
Features''~\citep{hensman2018variational} and ``Equispaced Fourier Gaussian
Process'' frameworks \citep{greengard2022equispaced}. Another close analog to
our approach is the method of \cite{im2007semiparametric}, which uses spline
bases to flexibly capture spectral densities at low frequencies and algebraic
tails at higher frequencies.

However, none of the above approaches provides a high order accurate method of
kernel evaluation which is compatible with performing maximum likelihood
estimation from a general spectral density $S$ with fully irregularly sampled
data. In this work, we demonstrate that panel Gaussian integration of the
Fourier integral allows the accurate and efficient computation of covariances
$K(r)$ and their derivatives $\pder{}{\theta_j}K(r)$ from \textit{any}
continuous, integrable spectral density $S$. This in turn facilitates
gradient-based maximum likelihood estimation directly on any parameterization of
the spectral density. By taking advantage of modern nonuniform fast Fourier
transform (NUFFT)~\citep{dutt1993fast,barnett2019parallel} and automatic
differentiation (AD)~\citep{griewank2008evaluating} methods, along with careful
analysis of the tail behavior of the spectral density, we can compute $K(r)$ and
its derivatives with respect to parameters to $\epsilon = \texttt{1e-12}$
accuracy even for slowly decaying spectral densities at millions of
inter-observation distances $r$ in seconds on a standard laptop. Our free and
open source \texttt{Julia} code is available at
\ifbool{anon}{\texttt{https://github.com/[author]/[package]}}{\texttt{https://github.com/pbeckman/SpectralKernels.jl}}.

While the possibilities for functional forms of spectral densities are endless
and the machinery described here is generally applicable, in this work we study
in detail the simple extension given by incorporating an integrable singularity
into a bounded spectral density $S(\omega)$, resulting in the new model
$|\omg|^{-\alpha} S(\omg)$. Such extensions correspond to ``long memory"
processes, for which we derive some theoretical properties, overcome numerical
challenges associated with evaluating their covariance functions, and
demonstrate their practical value by fitting a singular model to Doppler LiDAR
wind velocity profiles.

%% file: src/method.tex
We are concerned here only with real-valued covariance functions $K$, and
therefore assume $S$ is an even function. This results in the simplification
\begin{equation} \label{eq:real-fourier-integral}
    K(r) = 2\int_0^\infty S(\omega) \cos(2\pi\omega r) \, d\omega.
\end{equation} 
In order to evaluate $K(r)$ by directly computing
(\ref{eq:real-fourier-integral}), one must choose a quadrature rule. As the
spectral density $S$ is assumed to be integrable, it must decay sufficiently
fast for large $\omega$, so that one can truncate the infinite interval $[0,
\infty)$ at some point $b$ and integrate only on the finite interval $[0,b]$.
This gives
\begin{equation}
    K(r) 
    \approx 2\int_0^b S(\omega) \cos(2\pi\omega r) \, d\omega
    \approx 2\sum_{j=1}^m \gamma_j S(\omega_j) \cos(2\pi\omega_j r),
\end{equation}
where $\omega_j$ and $\gamma_j$ are the nodes and weights of the chosen
quadrature rule. The simplest choice of quadrature is the trapezoidal rule,
which uses equispaced points $\omega_j = (j-1)h$ with weights $\gamma_1 =
\gamma_m = \frac{h}{2}$ and $\gamma_j = h$ for $j=2,\dots,m-1$, where the grid
spacing is $h := \frac{b}{m-1}$. The aliasing and truncation errors when using
the trapezoidal rule are treated in detail for the squared exponential and
Mat\'ern models in~\cite{barnett2023uniform}.

For known spectral densities $S$ with sufficiently fast decay, this can be a
highly accurate quadrature. However, it has two limitations. First, for more
flexible and complex spectral densities $S$, the analysis used
in~\cite{barnett2023uniform} to choose the grid spacing $h$ necessary to resolve
$S$ becomes difficult, and must be done for each new $S$. Second, for small $r$
we may need to take both a small $h$ to control the quadrature error, as well as
a large $b$ to control the truncation error when integrating slowly decaying
spectral densities $S$ such as the commonly used Mat\'ern
model~\cite{stein1999interpolation}
\begin{equation}
    S_{\bth}(\omega) = \phi^2 (\rho^2 + \omega^2)^{-\nu - \frac{1}{2}}, \quad \bth := \{\phi, \rho, \nu\}
\end{equation}
with small values of the smoothness parameter $\nu$, e.g. $\nu = 1/2$. This can
result in a number of quadrature nodes $m$ which is prohibitively large from a
computational standpoint. In the remainder of this section, we demonstrate that
panel Gaussian integration of the Fourier integral allows us to overcome both of
these limitations.

\subsection{Panel integration of the Fourier integral}
\label{sec:panel-integration}

If one wishes to integrate (\ref{eq:real-fourier-integral}) for a large range of
$r$ with uniform accuracy, using a dense quadrature rule for $\omega \in [0,b]$
for some large $b$ might appear to be an unavoidable cost. However, as we
demonstrate numerically in this section and theoretically in
Section~\ref{sec:truncation-error}, truncation error in $K(r)$ is concentrated
near the origin in $r$-space. Intuitively, this is because the high frequency
information contained in the tails of the spectral density $S$ has a much
greater impact on covariances between observations which are close together.
This intuition will be made precise shortly by
Theorem~\ref{thm:truncation-error}.

The localization of error near $r = 0$ suggests that we can compute the Fourier
integral on the interval $\omega \in [0, b]$, then iteratively add the
contribution of the Fourier integral on successive intervals in $\omega$-space
for \textit{only} those $r$'s nearest the origin in $r$-space. We will refer to
each interval $[a, b]$ in $\omega$-space as a \textit{panel}. Define the exact
and discretized panel integrals
\begin{align}
    I_{[a,b]}(r) 
    &:= \int_a^b S(\omega) \cos(2\pi\omega r) \, d\omega \\
    \tilde{I}_{[a,b]}^{(m)}(r) 
    &:= \sum_{j=1}^m \gamma_j S(\omega_j) \cos(2\pi\omega_j r) \label{eq:panel-integral-quadrature}
\end{align}
where $\omega_j$ and $\gamma_j$ are nodes and weights of an $m$-point quadrature
rule on $[a,b]$. The error 
\begin{equation}
    E_{[a,b]}^{(m)}(r) := \abs{I_{[a,b]}(r) - \tilde{I}_{[a,b]}^{(m)}(r)}
\end{equation}
in the $m$-point trapezoidal rule with grid spacing $h := \frac{b-a}{m-1}$ is
only $\bO(h^2)$ when the integrand is non-periodic, as is the case in this panel
integral setting. However, the $m$-point Gauss-Legendre rule, which integrates
polynomials of degree $2m-1$ exactly, is generally much more accurate for
smooth, non-periodic functions. Gauss-Legendre quadrature is thus amenable to
panel integration and adaptivity, which allows us to accurately discretize
Fourier integrals for any continuous, integrable spectral density. Adaptive
quadrature will be treated in greater detail in
Section~\ref{sec:quadrature-error}.

By the Nyquist-Shannon sampling theorem, one requires at least $m =
2r\cdot(b-a)$ values to completely determine a function $f(\omega)$ with
bandlimit $r$ on $\omega \in [a,b]$. Therefore, for a desired accuracy $\delta$,
a given $m$, and sufficiently smooth $S$, we expect an $\bO(m)$-node
Gauss-Legendre rule to compute the Fourier integral with integrand $f(\omega) =
S(\omega) \cos(2\pi\omega r)$ to within accuracy $\delta$ on any interval of
length $b-a = \frac{m}{2r}$. We use this heuristic to choose the next panel
$[a,b]$ in $\omega$-space, where $r$ is taken to be the largest distance for
which $K(r)$ has not yet converged. Then all $r' < r$ result in less oscillatory
integrands, which are therefore also accurately resolved by the $m$-point rule.

It is worth noting that this Nyquist-informed heuristic for panel selection is
not tight in either direction. If $S$ contains sharp features, then the
integrand $f(\omega) = S(\omega) \cos(2\pi\omega r)$ will not have bandlimit
$r$, and thus more quadrature nodes may be required to achieve the desired
accuracy $\delta$. Conversely, if $S$ is very smooth, then the effect of
aliasing may be below the desired accuracy $\delta$, and thus fewer quadrature
nodes may still yield a adequately accurate result despite not fully resolving
the integrand. However, for sufficiently large $m$ we find that this
Nyquist-based choice typically computes the Fourier integral to double precision
with neither significant redundant oversampling nor the need for further
refinement where $S$ is smooth.

The novelty and efficiency of our approach stems from the two related mechanisms
discussed above. First, we can reduce the number of $r$ values to be computed
after adding each panel Fourier integral, as the largest $r$'s have converged.
This reduces the number of points $n$ at which we must evaluate the sum
(\ref{eq:panel-integral-quadrature}). Second, because $\cos(2\pi\omega r)$ is
less oscillatory in $\omega$ for smaller $r$, as the largest $r$'s converge we
can take increasingly large panels while keeping the number of oscillations per
panels constant. Therefore we use the same $m$-point quadrature rule to
accurately resolve the integrand. This can significantly reduce the total number
of quadrature nodes used to evaluate the Fourier integral when compared to a
more uniform quadrature scheme. Figure~\ref{fig:panel-integration} provides a
visual example of this panel growth as $K(r)$ is resolved for the largest $r$'s
which reduces the highest remaining Nyquist frequency. In particular, note that
the spacing between quadrature nodes increases by almost two orders of magnitude
between the first and fourth panels, corresponding to a proportional reduction
in computational effort relative to using, for example, the trapezoidal rule to
integrate the same interval in $\omega$-space. This trend only continues as we
progress in $\omega$-space, adaptively generating a highly non-uniform
quadrature rule with increasingly sparse nodes as $\omega$ increases. The
resulting quadrature provides orders-of-magnitude speedups over alternatives, as
we demonstrate in Section \ref{sec:dense-cov-error}.

\ifbool{figs}{
\begin{figure}
    \centering
    \includegraphics[width=\textwidth]{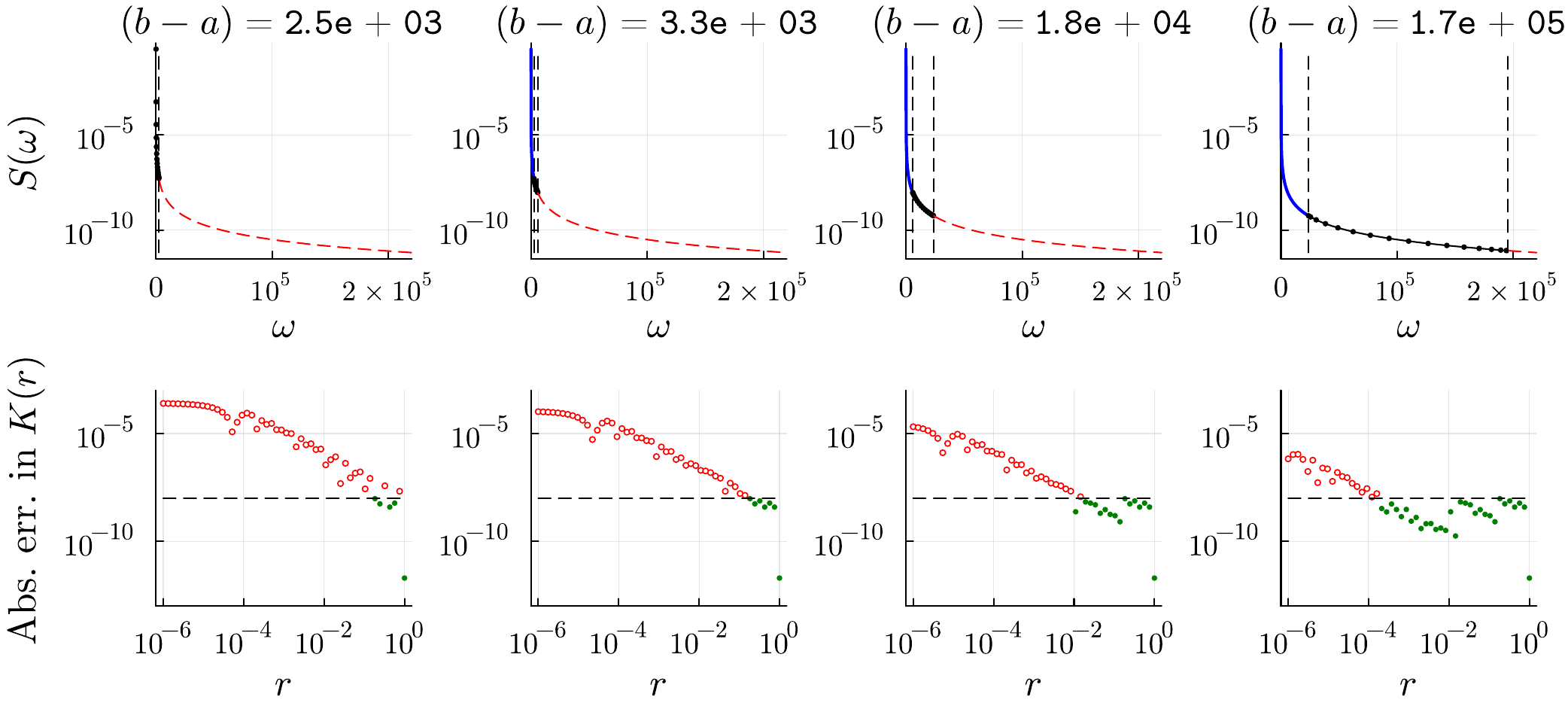} \\
    \caption{Panel integration of a Mat\'ern spectral density with
    $\nu=\frac{1}{2}, \rho=1$, and $\phi$ chosen so that $K(0) = 1$. The top row
    shows various panels of the spectral density being integrated, and each
    corresponding plot in the bottom row shows the absolute truncation error in
    the kernel after the panel integral above has been added. A small subset of
    the quadrature nodes are shown in each panel. A tolerance of $\epsilon =
    \texttt{1e-8}$ and $m=5{,}000$ nodes per panel are used. }
    \label{fig:panel-integration}
\end{figure}
}{}

Computing the sum (\ref{eq:panel-integral-quadrature}) directly for $n$
distances with an $m$-point quadrature rule has $\bO(nm)$ complexity for each
panel. This ostensibly introduces a major tension in the computation of these
panel integrals: higher-order quadrature rules enable larger panels and faster
convergence, but particularly for large data sizes they lead to enormously
burdensome computations if done directly. In this next section, we discuss how
the NUFFT can be used to relieve this tension by reducing the cost of each panel
integral to $\bO(m + n \log n)$ --- a speedup which is imperative to making
Fourier integration computationally tractable.

\begin{remark}
    The idea of using the same number of quadrature nodes $m$ to discretize
    Fourier integrals for which the space-frequency product $r\cdot(b-a)$ is
    constant is reminiscent of the complementary low-rank property used in
    butterfly algorithms~\cite{oneil2010algorithm,li2017interpolative}. One
    could view the present work as a butterfly-like algorithm for which the
    error control is designed specifically for the problem of adaptively
    computing continuous Fourier transforms of slowly decaying functions.
\end{remark}

\subsection{Acceleration with the nonuniform fast Fourier transform}

As each panel Fourier integral (\ref{eq:panel-integral-quadrature}) is a sum of
cosines with nonuniform frequencies $\omega_1,\dots,\omega_m$ evaluated at
nonuniform distances $r_1,\dots,r_n$, it can be computed as the real part of the
exponential sum
\begin{equation} \label{eq:nufft-type3-sum}
    f_k = \sum_{j=1}^m \gamma_j S(\omega_j) \expp{2\pi i \omega_j r_k} \quad \qquad k=1,\dots,n
\end{equation}
so that $\tilde{I}_{[a,b]}^{(m)}(r_k) = \Re(f_k)$. Equivalently, the computation
of the Fourier integral for each panel $[a,b]$ in $\omega$-space at all
unconverged distances $r$ can be viewed as a matrix-vector product of a dense $n
\times m$ nonuniform discrete Fourier matrix with a vector of $m$ evaluations of
the spectral density along with appropriate quadrature weights given by
\begin{equation} \label{eq:nufft-type3-matvec}
    \renewcommand*{\arraystretch}{1.5} 
    \text{$n$ distances}\left\{\vphantom{\mat{1\\2\\3}}\right.
    \underbrace{
    \begin{bmatrix}
    \expp{2 \pi i \omg_1 r_1} & \cdots & \expp{2 \pi i \omg_m r_1} \\
    \vdots & \ddots & \vdots \\
    \expp{2 \pi i \omg_1 r_n} & \cdots & \expp{2
    \pi i \omg_m r_n} \\
    \end{bmatrix}}_{\displaystyle \text{$m$ quadrature nodes}}
    \begin{bmatrix}
      \gamma_1 S(\omg_1) \\
      \vdots \\
      \gamma_m S(\omg_m) 
    \end{bmatrix}
    =
    \begin{bmatrix} 
        f_1 \\
        \vdots \\
        f_k
    \end{bmatrix}.
\end{equation}
Computing this dense matrix-vector product directly has a cost of $\bO(nm)$.
However, the sum (\ref{eq:nufft-type3-sum}) and the matrix-vector product
(\ref{eq:nufft-type3-matvec}) are equivalent views of exactly a ``type 3''
nonuniform to nonuniform discrete Fourier transform, which can be evaluated to
accuracy $\delta$ in $\bO(m + n\log n)$ complexity using the NUFFT. Most
existing NUFFT algorithms work by convolving the input data with some
``spreading'' function, performing an equispaced FFT on a fine grid, and
deconvolving to obtain values at the desired output locations. The development
of efficient NUFFT libraries with tunable accuracy guarantees has been the
subject of significant research in the past few decades. See for
example~\cite{barnett2019parallel,greengard2007fast,dutt1993fast}.

In the present context of computing pointwise covariances from the spectral
density, the improved scaling of the NUFFT when compared to direct summation is
essential to facilitating the use of larger, more accurate quadrature rules to
compute the values of each panel integral at a greater number of distances. The
dramatic computational impact of the NUFFT will be studied in greater detail in
Section~\ref{sec:dense-cov-error}, where we observe orders of magnitude speedup
for practical data sizes.

\begin{remark}
    The number of points $m$ in each NUFFT can be tuned to improve performance.
    Generally, using large NUFFTs helps to amortize setup costs and take
    advantage of multi-threading. However, ill-conditioning and round-off errors
    limit the accuracy of the NUFFT to about \texttt{1e-9} for large
    inputs~\cite[Remark 9]{barnett2019parallel}. So if higher accuracy is
    needed, smaller NUFFTs must be used at the expense of a larger pre-factor.
\end{remark}

\subsection{Error estimation} \label{sec:error-estimation}

As we saw in Section~\ref{sec:panel-integration}, the key idea behind fast panel
integration of Fourier integrals is to use panels of increasing size in
$\omega$-space. This is made possible by choosing a working tolerance $\delta$
and sequentially integrating panels in $\omega$-space for \textit{only} those
distances $r$ at which $K(r)$ has not yet been resolved to within $\delta$
accuracy. Determining when $K(r)$ is adequately resolved for each $r$ requires a
careful analysis of multiple sources of error, and naive stopping criteria often
result in the loss of several digits of accuracy in the computed integral.

Recall that for any valid covariance function $K$, one has that $K(0) \geq
\abs{K(r)}$ for all $r > 0$. Then for any set of locations $x_1,\dots,x_n$,
computing each entry in an approximate covariance matrix $\widetilde{\bS}_{ij}
:= \widetilde{K}(x_i - x_j)$  such that the pointwise error relative to $K(0)$
is controlled $ |K(r) - \widetilde{K}(r)| \ / \ K(0) < \epsilon 
$
for all $r$, consequently bounds the relative max-norm error in
$\widetilde{\bS}$
\begin{equation}
    \frac{\shortnorm[\text{max}]{\bS - \widetilde{\bS}}}{\norm[\text{max}]{\bS}}
    = \frac{\displaystyle\max_{1 \leq i,j \leq n} \abs{K(x_i - x_j) - \widetilde{K}(x_i - x_j)}}{K(0)} 
    < \epsilon.
\end{equation}
By the equivalence of norms, controlling the relative max-norm error also
controls the relative Frobenius and spectral norm errors, up to constants which
may depend on $n$. As our real goal is to evaluate the log-likelihood, which,
for fixed data $\by$ is a smooth function of the covariance matrix $\bS$,
controlling relative norm errors in $\widetilde{\bS}$ is perhaps the most
relevant metric.

We emphasize that the pointwise error relative to $K(0)$ is equivalent to
neither the relative \textit{nor} the absolute pointwise error in each $K(r)$.
If $K(r) = \texttt{1e-12}$, then \textit{relative} error $|K(r) -
\widetilde{K}(r)| / \abs{K(r)} < \texttt{1e-8}$ would require cancellation in
the sum of panel integrals to 20 digits, which is impossible in general in
double precision. Conversely, if $K(r) = \texttt{1e12}$, then \textit{absolute}
error $|K(r) - \widetilde{K}(r)| < \texttt{1e-8}$ would require 20 correct
digits, which is again impossible in double precision. These are standard and
well-known limitations of adaptive integration in finite precision arithmetic,
and neither bounding the relative nor absolute pointwise error by $\epsilon$ are
necessary conditions for bounding the resulting error in log-likelihood
evaluation by $\epsilon$. There are two sources of error at play here which we
seek to bound above by $\delta$ --- quadrature error and truncation error ---
which we now treat individually.

\subsubsection{Quadrature error} \label{sec:quadrature-error}

To control the quadrature error, we use a straightforward adaptive 
approach. As is standard practice in adaptive integration, we estimate the error
in the discretized integral by comparing the result to that obtained by a higher
order quadrature rule. For this purpose we use a $2m$-point Gauss-Legendre rule.
The error in the $m$-point rule on $[a,b]$ is then approximated as
\begin{equation}
    \widetilde{E}_{[a,b]}^{(m)}(r)
    := \abs{\tilde{I}_{[a,b]}^{(2m)}(r) - \tilde{I}_{[a,b]}^{(m)}(r)} 
    \approx \abs{I_{[a,b]}(r) - \tilde{I}_{[a,b]}^{(m)}(r)}.
\end{equation}
For a given panel, if $\widetilde{E}_{[a,b]}^{(m)}(r) > \delta$ for any
unconverged distance $r$, we divide $[a,b]$ in two and repeat this procedure
separately on $[a, (a+b)/2]$ and $[(a+b)/2, b]$. We deem these subpanels
converged when both $\widetilde{E}_{[a,(a+b)/2]}^{(m)}(r) < \delta/2$ and
$\widetilde{E}_{[(a+b)/2, b]}^{(m)}(r) < \delta/2$, so that the total error on
$[a,b]$ remains bounded by $\delta$ for all unconverged $r$'s. If the error
remains large for any $r$ in one or both of these subpanels, we continue this
dyadic refinement and proportional tolerance reduction until we obtain an
approximation $\tilde{I}_{[a,b]}(r)$ as a sum of subpanel integrals such that
the total quadrature error is bounded by $\delta$. This is again standard
practice for adaptive integration, and can be performed to high accuracy. See
for example~\cite{gonnet2012review}.

\subsubsection{Truncation error} \label{sec:truncation-error}

Controlling the truncation error, given by 
\begin{equation}
    E_{\text{trunc}}(b,r) := \int_b^\infty S(\omega) \cos(2\pi\omega r) \,
    d\omega,
\end{equation}
is a more subtle issue. As discussed in Section~\ref{sec:panel-integration}, we
iteratively integrate panels from zero to infinity in spectral space. The
remaining question is how to determine for each $r$ when we have integrated a
sufficient interval $[0,b]$ in spectral space so that
$\abs{E_{\text{trunc}}(b,r)} < \delta$ and we can cease adding new panels.

One could check that the contribution of the current panel is less than the
tolerance $\delta$, that is $\tilde{I}^{(m)}_{[a,b]}(r) < \delta$. This is a
necessary but \textit{not} a sufficient condition for $E_{\text{trunc}}(b,r) <
\delta$. In practice, for spectral densities which decay exponentially this
condition results in a negligible loss of accuracy. However, for spectral
densities with slow algebraic decay, terminating integration when
$\tilde{I}^{(m)}_{[a,b]}(r)  < \delta$ may result in errors significantly
greater than $\delta$, as one would be truncating many panels whose
contributions would each be $\bO(\delta)$. 

Therefore, for better error control, we consider the exponent $\beta$ and
constant $c$ in the tails of the spectral density such that $S(\omega) \sim
c\omega^{-\beta}$ as $r \to \infty$. For many spectral densities $c$ and $\beta$
can be derived analytically, often by straightforward means. Otherwise, these
coefficients can be estimated using linear least squares in log space. With
these values, we can compute the truncation error analytically for the resulting
power law tail, which gives
\begin{align} \label{eq:truncation-error}
    \abs{E_{\text{trunc}}(b,r)}
    &\approx \abs{\int_b^\infty c\omega^{-\beta} \cos(2\pi\omega r) \, d\omega} \\
    &= -c (2\pi r)^{\beta-1} \Re\Big((-i)^{\beta-1} \Gamma(-\beta+1, -2\pi
    ibr)\Big),
\end{align}
where $\Gamma(s, z) := \int_z^\infty t^{s-1} \expp{-t} \, dt,$ is the upper
incomplete Gamma function \citep{olver2010nist}. There exist a number of
libraries to numerically evaluate this special function. However, it is helpful
to have a simple, tight, and easily invertible algebraic upper bound on this
truncation error. We now derive such a bound  as a consequence of the following
lemma, whose proof is given in the appendix.

\begin{lemma} \label{lem:inc-gamma} For any $s, y > 0$,
    \begin{equation}
        \abs{\Gamma(-s, iy)} \leq \min\left(y^{-s-1}, \frac{y^{-s}}{s}\right).
    \end{equation}
\end{lemma}

\begin{theorem} \label{thm:truncation-error} For any $\beta > 1$ and any $b, r >
    0$,
    \begin{equation} \label{eq:truncation-estimate}
        \abs{\int_b^\infty \omega^{-\beta} \expp{-2\pi i\omega r} \, d\omega}
        \leq \min\left(
            \frac{1}{\beta-1} b^{-\beta+1},
            \frac{1}{2\pi r} b^{-\beta}
        \right),
    \end{equation}
    where the first term gives a tighter bound when $br \leq
    \frac{\beta-1}{2\pi}$, and the second term otherwise.
\end{theorem}
\begin{proof}
    Taking the change of variables $t = 2\pi i\omega r$, we obtain
    \begin{align}
        \abs{\int_b^\infty \omega^{-\beta} \expp{-2\pi i\omega r} \, d\omega}
        &= \abs{(2\pi i r)^{\beta-1} \int_{2\pi ibr}^\infty t^d \expp{-t} \, dt}
        = (2\pi r)^{\beta-1} \big|\Gamma(-\beta+1, 2\pi ibr)\big|. \nonumber
    \end{align}
    Applying Lemma~\ref{lem:inc-gamma} gives the desired result.
\end{proof}

This result indicates that when the product $br$ is small (for close together
observations or limited integration domains in spectral space), the truncation
error in the Fourier integral behaves like the truncation error in the
non-oscillatory integral. But when $br$ is large (for far apart observations or
large integration domains in spectral space), the truncation error decays faster
by an additional power of $b$, with a constant that decreases for larger $r$.
This makes precise the intuition of Figure~\ref{fig:panel-integration} that for
a fixed integration domain, the truncation error for larger distances $r$ is
lower.

One can therefore use the analytic formula (\ref{eq:truncation-error}) or the
simpler algebraic upper bound (\ref{eq:truncation-estimate}) to control the
truncation error. In conjunction with the adaptive integration and quadrature
error estimation described in Section~\ref{sec:quadrature-error}, we obtain
accurate estimates of the total error in the computed Fourier integral.

\ifbool{figs}{
\begin{figure}
    \hspace{0.16\textwidth}
    $\abs{\Gamma(-s, iy)}, \ s=1$ 
    \hspace{0.12\textwidth} 
    $\int_b^\infty \omega^{-\beta} \cos(2\pi\omega r) \dif \omega, \ \beta = 2 $
    \\
    \centering
    \includegraphics[width=0.8\textwidth]{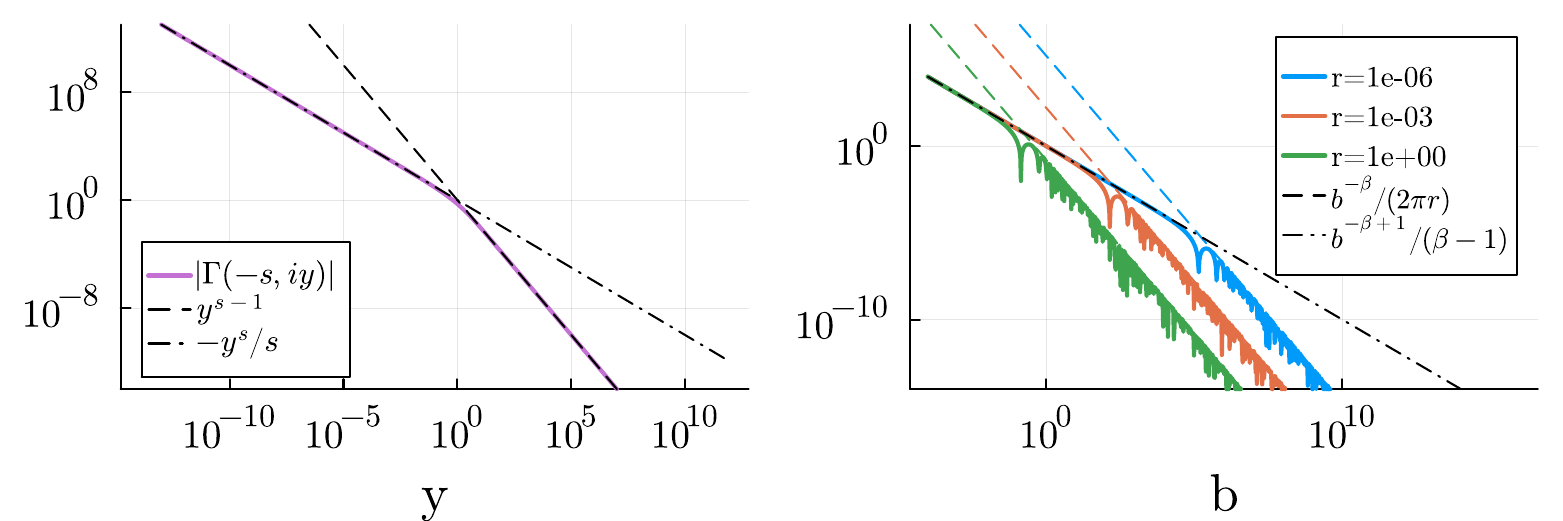}
    \caption{The incomplete Gamma function bound from Lemma \ref{lem:inc-gamma}
    (left) and corresponding bound on the power law truncation error given by
    Theorem \ref{thm:truncation-error} for various $r$ (right).}
    \label{fig:truncation-bound}
\end{figure}
}{}

\subsection{Power law singularities at the origin}

Long-memory Gaussian processes can be characterized by having an integrable
singularity in their spectral density $S$ at the origin. Such models typically
lack closed form expressions for the resulting kernel $K$, and even when such
expressions are available, they are often difficult to compute numerically. In
this work, we will focus on a modification of the Mat\'ern family that we will
call a ``singular" Mat\'ern model, which was first proposed in
\citep{porcu2012some} and corresponds to a spectral density given by
\begin{equation} \label{eq:singularmatern}
  S_{\bth}(\omg) = \phi^2 |\omg|^{-\alpha} (\rho^2 + \omg^2)^{-\nu - \frac{1}{2}}, \quad \bth := \{\phi, \alpha, \rho, \nu\}
\end{equation}
with $ 0 \leq \alpha < 1$. As will be discussed in detail in the next section,
the corresponding covariance function does technically have a closed form
representation, but it is exceptionally challenging to evaluate numerically.

In contrast, such spectral densities with power law singularities are treated
easily by the panel Gaussian quadrature framework presented here. While directly
applying a Gauss-Legendre quadrature to the integral
\begin{equation} \label{eq:singular-fourier-integral}
    K(r) = 2\int_0^b \omega^{-\alpha} S(\omega) \cos(2\pi\omega r) \, d\omega
\end{equation}
results in low accuracy, Gauss-Jacobi quadratures are designed specifically to
treat the singularity $\omega^{-\alpha}$ accurately, and can be computed quickly
and accurately~\citep{glaser2007fast, hale2013fast}. Simply using a Gauss-Jacobi
rule on the first panel followed by Gauss-Legendre rules on all remaining panels
allows accurate computation of singular Fourier integrals of this form. For
panels that do not contain the origin, the $\omega^{-\alpha}$ term may safely be
included in the Fourier integrand, and the standard Gauss-Legendre rule gives
high accuracy.

\subsection{Computing derivatives of the kernel}

In order to take advantage of gradient-based optimizers for maximum likelihood
estimation, one must compute derivatives $\pder{}{\theta_j}K(r)$. Assuming that
the parametric family $S_{\bth}$ is sufficiently well-behaved as to allow
exchanging differentiation and integration, we have
\begin{equation} \label{eq:dkernel}
    \pder{K_{\bth}(r)}{\theta_j}
    = 2\int_0^\infty \omega^{-\alpha} \pder{S(\omega)}{\theta_j} \cos(2\pi\omega r) \, d\omega,
\end{equation}
which can be computed using the same framework that is used to evaluate
$K_{\bth}$ itself.

While differentiating even standard covariance functions such as the Mat\'ern
with respect to kernel parameters can be challenging \citep{geoga2023fitting},
most common spectral densities are very simple to differentiate. The computation
of the partial derivatives $\pder{}{\theta_j}S_{\bth}(\omg)$ can be done easily
using automatic differentiation (AD) \citep{griewank2008evaluating}, which
operates at the code level to programmatically generate derivatives of a given
function. As a result, end users can simply write any parametric spectral
density $S_{\bth}$ that they would like, and the derivatives of $K_{\bth}$ will
be obtained automatically using our software.

This ease of differentiation can be extended to singular spectral densities. To
compute the derivative of the kernel $K$ with respect to the singularity
parameter $\alpha$, given by
\begin{equation}
    \pder{K(r)}{\alpha} 
    = -2\int_0^\infty \omega^{-\alpha} \log(\omega) S(\omega) \cos(2\pi\omega r) \, d\omega,
\end{equation}
we require a method for accurately integrating the
$\omega^{-\alpha}\log(\omega)$ singularity on the panel $[0,b]$ containing the
origin. Ignoring the $\log(\omega)$ singularity and applying a Gauss-Jacobi rule
results in low accuracy, especially as $\alpha$ approaches 1 and the weights
become relatively large and positive near zero. However, if we apply integration
by parts
\begin{align}
    (1-\alpha) \int_0^b \omega^{-\alpha} \log(\omega) S(\omega) \cos(2\pi\omega r) \, d\omega
    &= b^{1-\alpha}\log(b)S(b) \cos(2\pi b r) \nonumber \\
    &\hspace{-0.1\textwidth}- \int_0^b \omega^{-\alpha} \Big( S(\omega) +  \omega\log(\omega) S'(\omega) \Big) \cos(2\pi\omega r) \, d\omega \\
    &\hspace{-0.1\textwidth}+ 2\pi r \int_0^b \omega^{1-\alpha} \log(\omega) S(\omega) \sin(2\pi\omega r) \, d\omega, \nonumber
\end{align}
we see that the resulting integrals involve only the power law singularity
$\omega^{-\alpha}$, and that the log singularity has been removed. As $S$ is a
closed form function provided by the user, $S'(\omega)$ can be obtained
analytically or using AD, and we can evaluate the above expression using the
same Gauss-Jacobi rule that is employed to compute kernel values. This requires
only two NUFFTs, and avoids any additional cost to compute specialized
quadratures for the $\omega^{-\alpha}\log(\omega)$ singularity. For panels which
do not contain the origin, the $\omega^{-\alpha}\log\omega$ term can be included
in the integrand, and the Gauss-Legendre rule gives high accuracy as before.
Using this strategy and incorporating it into a custom rule for the AD engine
again means that this derivative can be obtained entirely programmatically
without any end-user intervention.

%% file: src/results.tex
Before employing the adaptive integration method described above to compute
kernel values in a maximum likelihood estimation context with real data, we
provide several demonstrations. First, we give examples of statistically
interesting models which can be written easily in the spectral domain but for
which no closed form expression exists for the corresponding covariance
function. Next, we show that our method can accurately evaluate the singular
Mat\'ern covariance function even when existing alternative numerical methods
fail. We close with a runtime comparison which illustrates that our adaptive
NUFFT-accelerated Gaussian quadrature scheme is necessary to efficiently obtain
high accuracy kernel evaluations.

The online supplement to this work contains several additional numerical
demonstrations and examples. In particular, it provides a demonstration of the
sharpness of the error control described in Section~\ref{sec:error-estimation}
applied to the standard Mat\'ern model, as well as further numerical experiments
involving the singular Mat\'ern model that validate the correctness of our
method and compare it to alternatives.

\subsection{Designing new spectral densities}
\label{sec:interesting}

As a motivating example to illustrate how easily practitioners can write models
in the spectral domain, we study a generalization of the standard Mat\'ern model
$K(r) \propto r^{\nu} \mathcal{K}_{\nu}(r)$. Two known limitations of the
Mat\'ern model are its exponential decay as $r \to \infty$ for any finite $\nu$,
and its inability to take on negative values. While loosening these restrictions
in ``kernel space" is challenging, it is trivial to write spectral densities
whose corresponding kernels move beyond these limitations. Consider, for
example, the model
\begin{equation} \label{eq:generalizedmatern}
  S_{\bth}(\omg)
  =
  \phi^2
    \big(\lam + (1-\lam)\abs{\omg}^\gamma\big)
    \big(\rho^2 + \abs{\omg}^{\tau}\big)^{-\nu - \frac{1}{2}},
    \quad \bth := \{\phi, \lambda, \gamma, \rho, \tau, \nu\},
\end{equation}
with $\lambda \in [0,1]$, $\tau \in (0, 2]$, and the constraint $\tau(\nu + 1/2)
- \gamma > 1$ for integrability. In the case $\tau = 2$ this is precisely a
standard Mat\'ern model plus a fractional derivative of a standard Mat\'ern
model. For $\tau < 2$ the function $S_{\bth}$ will not be smooth at the origin,
so the corresponding kernel will decay more slowly than the exponential rate of
the standard Mat\'ern.  For small values of $\lambda$ and large enough values of
$\gamma$, the corresponding kernel can also take negative values. Figure
\ref{fig:example-genmatern} shows an example of this model. \ifbool{figs}{
\begin{figure}[!ht]
  \centering
  \includegraphics[width=\textwidth]{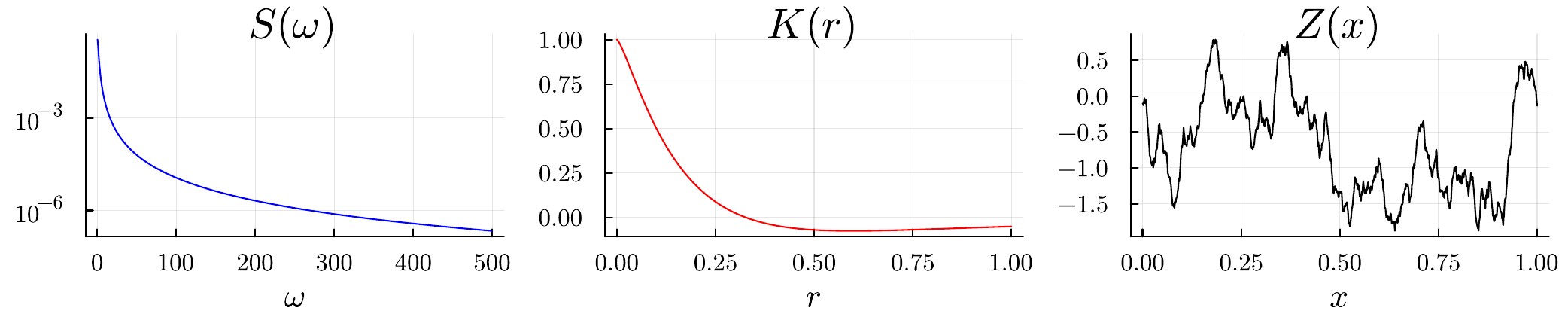}
  \caption{The generalized Mat\'ern spectral density
  (\ref{eq:generalizedmatern}) (left), its corresponding covariance function
  (center), and a sample path from the process (right).}
  \label{fig:example-genmatern}
\end{figure}
}{}

To further highlight the broad class of novel models which can be specified
through their spectral densities and fit using our method, we consider two more
examples. The first is an ``oscillatory" Mat\'ern spectral density given by
\begin{equation} \label{eq:oscillatory-matern} 
    S_{\bth}(\omega) = \phi^2
    (\rho^2 + \omega^2)^{-\nu-\frac{1}{2}} \Big(1 -
    \expp{-\lambda|\omega|}\sin(\gamma|\omega|)\Big), \quad \bth := \{\phi, \rho, \nu, \lambda, \gamma\}.
\end{equation} 
This model promotes oscillatory behavior through strong negative then positive
kernel values near the origin, while still maintaining full control over the
mean-square differentiability of the process through the parameter $\nu$.
\ifbool{figs}{
\begin{figure} 
    \centering
    \includegraphics[width=\textwidth]{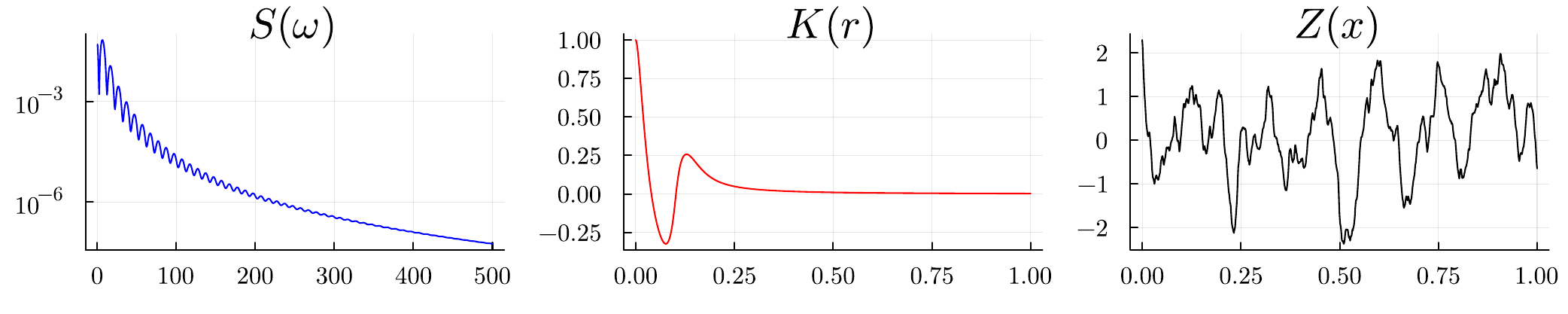}
    \includegraphics[width=\textwidth]{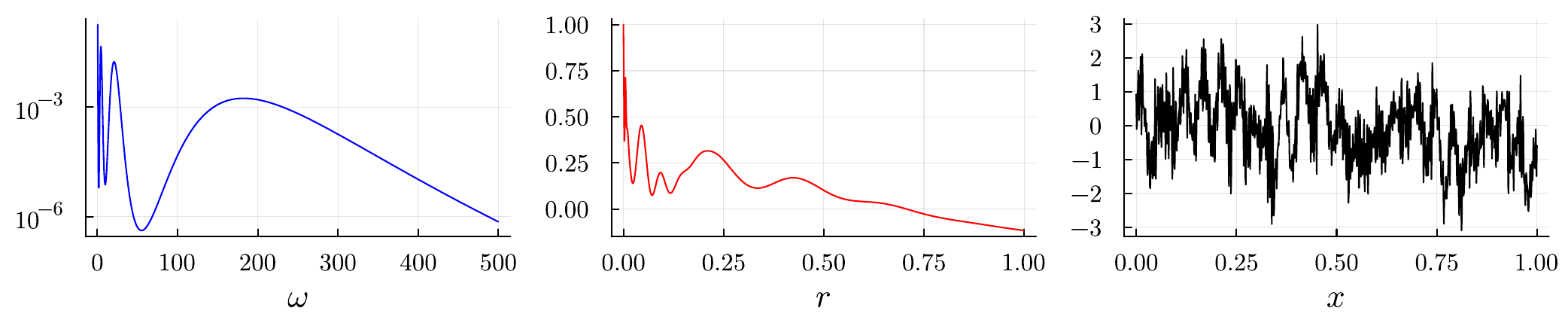}
    \caption{The top row shows the  ``oscillatory'' Mat\'ern spectral density
    (\ref{eq:oscillatory-matern}) (left), its corresponding covariance function
    (center), and a sample path from the process (right). The bottom row
    displays analogous quantities for the semi-parametric long-memory model
    (\ref{eq:chebyshev-exponential}).} \label{fig:example-sdfs} 
\end{figure}
}{}

The final model we consider here is a semi-parametric long-memory model given by 
\begin{equation} \label{eq:chebyshev-exponential} 
    S_{\bth}(\omega) =
    \phi^2 |\omega|^{-\alpha} \exp\left\{-\lambda|\omega| + \sum_{k=0}^K c_k
    T_k\left(\frac{|\omega|-\rho}{|\omega|+\rho}\right)\right\}, \quad \bth :=
    \{\phi, \alpha, \lambda, \rho, c_0, \dots, c_K\}, 
\end{equation} 
where $T_k$ is the Chebyshev polynomial of order $k$. This model is in the
spirit of semi-parametric ideas like the popular \emph{spectral mixture kernel}
\citep{wilson2013gaussian} in that it avoids directly specifying the functional
form of the spectral density. The singularity parameter here again gives the
model the flexibility to capture slowly decaying tails in the covariance
function, and the arbitrary number of orthogonal polynomials affecting
frequencies near the origin can yield flexible kernels. Both of these models are
shown in the spectral domain, covariance domain, and with an example sample path
in Figure \ref{fig:example-sdfs}.

\subsection{Singular Mat\'ern model} \label{sec:singularmatern}

As previously mentioned, the singular Mat\'ern model (\ref{eq:singularmatern})
is a useful tool for validating our approach and demonstrating its effectiveness
for spectral densities with origin singularities. The Fourier transform of
(\ref{eq:singularmatern}) is given by
\begin{align} \label{eq:singularmatern_kernel} 
    K(r) &= 2
    \int_{0}^{\infty} |\omg|^{-\alpha} \phi(\rho^2 + \omg^2)^{-\nu - 1/2} \cos(2 \pi
    \omg r) \dif \omg \\
    &= \frac{\phi}{\rho^{2\nu} \, (2 \pi r)^{-\alpha} \ \Gamma(\nu + \tfrac{1}{2})} \bigg[
        (2 \pi r \rho)^{-\alpha} \, \Gamma(\tfrac{2\nu + \alpha}{2})
    \Gamma(\tfrac{1-\alpha}{2}) \,_{1}F_{2}\Big(\tfrac{1-\alpha}{2}, \{\tfrac{1}{2},
    \tfrac{2-2\nu-\alpha}{2}\}, (\rho \pi r)^2\Big) \nonumber
     \\
    & \hspace{0.04\linewidth} + 2 (2\pi r\rho)^{2\nu} \cos(\tfrac{(2\nu + \alpha)\pi}{2})
    \Gamma(\tfrac{2\nu+1}{2}) \Gamma(-\alpha - 2\nu)  \,_{1}F_{2}\Big(\tfrac{2\nu+1}{2},
    \{\tfrac{1+2\nu+\alpha}{2}, \tfrac{2+2\nu+\alpha}{2}\}, (\rho \pi r)^2\Big) \bigg], \nonumber
\end{align}
where $\Gamma$ is the Gamma function and $_{1}F_{2}$ is a generalized
hypergeometric function \citep{olver2010nist}. Evaluating this $K(r)$ accurately
in double precision is extremely challenging, as the two terms inside the
brackets are rapidly growing numbers of opposite signs which approach each other
in absolute value and cancel as $r$ grows. In exact arithmetic this is not a
problem, but it poses a serious numerical issue in finite precision. Consider
the case where both of the above terms have absolute value of order
\texttt{1e20}, and a routine for evaluating any of the constituent special
functions incurs a numerical relative error of even \texttt{1e-15}. Such small
relative errors are inevitable, as double precision floating point numbers store
only roughly 16 relative digits. Then an addition which should result in exact
cancellation to zero could give a value of size $\texttt{1e-15} \times
\texttt{1e20} \approx \texttt{1e5}$ instead. Considering how quickly the inner
terms grow, one reaches this regime for small $r$ even with non-pathological
choices of $\phi$, $\rho$, and $\nu$. 

To illustrate the challenge of evaluating the singular Mat\'ern covariance
function (\ref{eq:singularmatern_kernel}), we compare three methods for
computing $K(r)$: one using a double-precision library for all special
functions, one using the extended precision mathematical library \texttt{Arb}
\citep{Johansson2017} for all special functions with $10{,}000$ bits of
precision, and our method. \ifbool{figs}{
\begin{figure}[!ht]
  \centering
  \includegraphics[width=\textwidth]{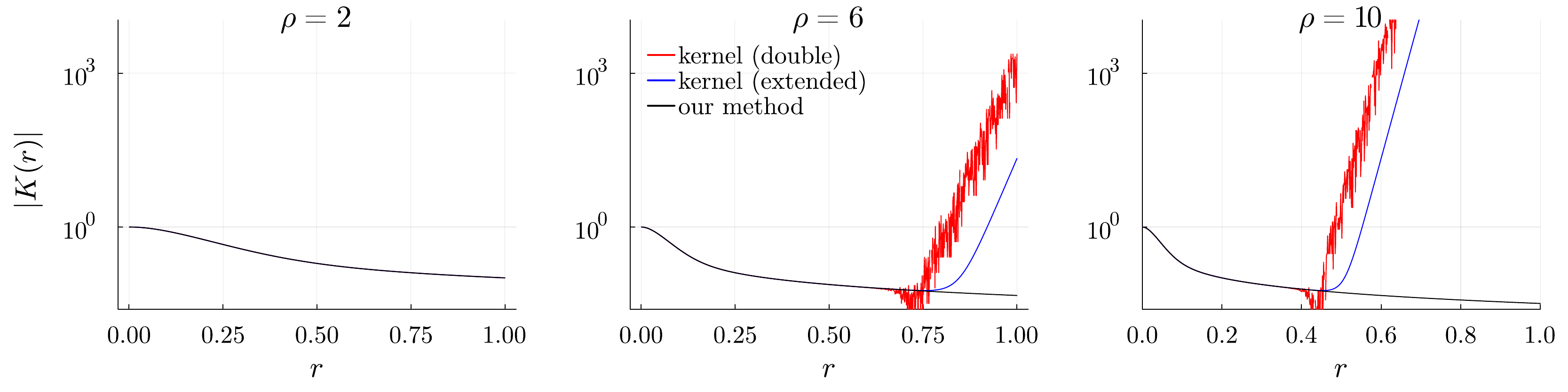}
  \caption{A comparison of the three methods for evaluating the covariance
  function given in Equation \ref{eq:singularmatern_kernel}: the
  double-precision direct kernel routine (red), the extended precision direct
  kernel routine (blue), and our Fourier quadrature routine (black).}
  \label{fig:singular_matern}
\end{figure}
}{} Figure \ref{fig:singular_matern} shows the evaluated kernel on a regular
grid of points in $[0,1]$ for a choice of $\nu=2.1$, $\alpha=0.3$, several
$\rho$ values, and in all cases a $\phi$ such that $K(0) = 1$, which is a ``best
case" numerical choice. For the case $\rho=2$, Figure \ref{fig:singular_matern}
shows good visual agreement between the three methods. But as $\rho$ is
increased, we can see that both direct evaluation methods clearly fail to
achieve even a single correct digit by $r=0.8$. For $\rho=10$, which is again
still not a pathological parameter choice, we see that both direct evaluation
methods are so affected by roundoff errors ruining cancellation that they
provide $K(1)$ to be on the order of $\texttt{1e14}$. As an additional
indication of correctness even when direct methods cannot be used for
validation, the online supplement provides summary tables that show that the
kernel values obtained with our quadrature-based method correctly provide
positive-definite matrices for all tested parameter values.

The difficulty of accurately evaluating $K(r)$ for large $r$ is especially
problematic because long-memory models are most commonly applied to data which
are strongly dependent on long time horizons --- and it is precisely these cases
in which evaluating the covariance function for large $r$ is most relevant. This
is because singular spectral densities can be used to build covariance functions
that decay exceptionally slowly. As the following theorem demonstrates, singular
spectral densities can provide covariance functions that are square-integrable
but not integrable, or for $\alpha > 1/2$ not even square-integrable.
\begin{theorem} \label{thm:decay} Let $S \in C^2(\R) \cap L^1(\R)$ be a bounded,
  symmetric, positive spectral density with $S', S'' \in L^1(\R)$. Then for all
  $\alpha \in [0,1)$
  \begin{equation*} 
    K_\alpha(r) 
    := 2\int_0^\infty |\omega|^{-\alpha} S(\omega) \cos(2 \pi \omega r) \dif \omega 
    \sim r^{-1+\alpha} \quad \text{as} \ r \to \infty.
  \end{equation*}
\end{theorem}
We note that the requirement $S \in C^2(\R)$ is far from a necessary condition,
and this asymptotic behavior is a very general phenomenon. For example
$S(\omega) = \expp{-\abs{\omega}}$ yields the covariance function 
\begin{align}
  K_\alpha(r)
  & = \frac{2}{\Gamma(1-\alpha)} \Big( 1 + (2\pi r)^2 \Big)^{-\frac{1}{2}(1-\alpha)} \cos\Big((1-\alpha) \tan^{-1}(2\pi\abs{r})\Big)
  \sim r^{-1+a}
\end{align}
as $r \to \infty$ despite the fact that $S$ is not even $C^1(\R)$. In such cases
where $S$ is monotonic on $[0,\infty)$, one can apply more general Tauberian
results from the theory of slowly varying functions to obtain the same
asymptotic decay \cite[Theorem 4.10.3]{bingham1989regular}.

\begin{remark}
  It is possible that one could use an expansion of the hypergeometric
  function in conjunction with relevant special function identities to derive
  an evaluation scheme for (\ref{eq:singularmatern_kernel}) which avoids the
  numerical instability of adding very large floats. However, doing so
  efficiently and accurately would represent a significant research endeavor,
  and our method can evaluate a broad class of such covariance functions
  without requiring further kernel-specific effort.
\end{remark}

\subsection{Performance for dense covariance matrix construction}
\label{sec:dense-cov-error}

As discussed in Section \ref{sec:error-estimation}, by controlling the pointwise
error relative to $K(0)$ we have bounded the relative max-norm error in any
covariance matrix with entries computed using our method. In
Figure~\ref{fig:digits}, we demonstrate agreement between the user-specified
tolerance $\epsilon$ and the relative error in $\bS$ in various matrix norms
with $N=1000$ random observation locations $x_1, \dots, x_N \iid
\text{Unif}([0,1])$ for a slowly-decaying singular Mat\'ern model with
$\nu=0.51, \rho=0.5$, and $\alpha = 0.1$, with $\phi$ chosen so that $K(0)=1$.
We also show analogous errors in the derivative matrices $\pder{}{\theta}\bS$
computed using AD for parameters $\theta \in \{\nu, \alpha\}$, for which the
derivatives are most numerically challenging. \ifbool{figs}{
\begin{figure}
    \centering
    \includegraphics[width=\textwidth]{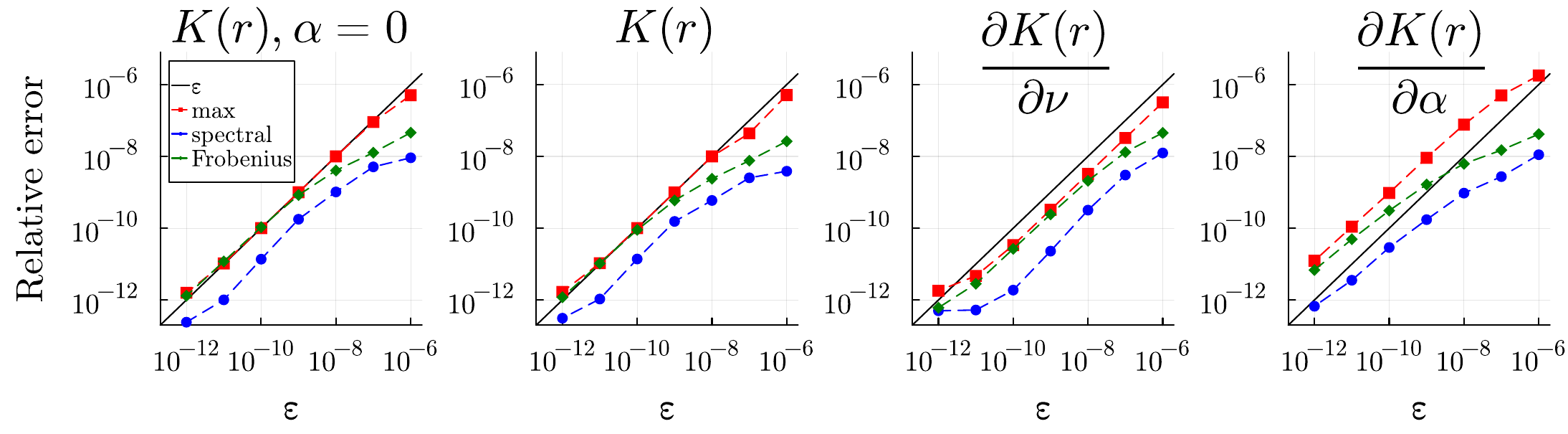}
    \caption{Comparison of the relative error $\shortnorm{\bA -
    \tilde{\bA}}/\norm{\bA}$ with the requested tolerance $\epsilon$ for a
    variety of matrix norms, where $\bA = \bS$ (first and second plots) or $\bA
    = \pder{}{\theta}\bS$ (third and fourth plots). In the first plot, $\alpha$
    is set to zero to test the non-singular case, and all other parameters
    remain unchanged.}
    \label{fig:digits}
\end{figure}
}{}

Having confirmed the accuracy of our scheme, we now study its computational
cost, and demonstrate that both the NUFFT and our Nyquist-based heuristic for
increasing panel length are necessary for computational efficiency. For various
$N$ ranging from 10 to 10{,}000, we form the dense covariance matrix for $N$
uniform random observation locations in $[0,1]$ as above and time three
approaches. First, we run our adaptive Gauss-Legendre method and time only the
evaluation of the necessary NUFFTs, excluding the computation of error
estimates. This provides a fair comparison with the trapezoidal rule, which is
non-adaptive. Second, we repeat this with direct Fourier sums in place of the
NUFFT. Finally, we use \cite[Corollary 6]{barnett2023uniform} to determine the
grid spacing $h$ and number of quadrature nodes $m$ in a trapezoidal rule
necessary to obtain each tolerance, and time the ``type 2'' NUFFT needed to
compute kernel values from this quadrature rule. Figure~\ref{fig:timing} shows
the resulting runtimes using an 8 core Apple M1 Pro CPU with 32GB of memory.

There are a two significant conclusions to be gathered from these results. The
most obvious is the paramount importance of the NUFFT. The difference between
$\bO(nm)$ and $\bO(m + n\log n)$ complexity for direct summation and the NUFFT
respectively gives an orders of magnitude speedup, without which one can afford
to compute only very few kernel values. The other conclusion is the necessity of
our Nyquist-based heuristic for obtaining high accuracy kernel values. For
tolerance $\epsilon = \texttt{1e-8}$, the trapezoidal rule requires both small
$h$ in order to resolve the spectral density near the origin, as well as large
$m$ to control the truncation error. This results in a very large NUFFT of size
$m \approx \texttt{1e8}$. The $\bO(m)$ ``spreading'' step in the NUFFT thus
becomes the dominant cost for all tested $n$, leading to a nearly constant cost
which is orders of magnitude slower than our adaptive quadrature for moderate
$n$. For $\epsilon = \texttt{1e-12}$, the necessary number of equispaced
trapezoidal nodes is $m \approx \texttt{1e12}$, and results in memory issues. As
our adaptive scheme uses panels with $m = 2^{16}$ nodes, the $\bO(m)$
``spreading'' step is only the bottleneck up to $n = \texttt{1e5}$ or so, after
which the $\bO(n\log n)$ equispaced FFT which is evaluated within the NUFFT
routine becomes the dominant cost, and we see quasilinear scaling with $n$.
Therefore our method still runs in seconds, even when computing the kernel at
approximately 50 million distances to 12 digit accuracy. \ifbool{figs}{
\begin{figure}
    \centering
    \includegraphics[width=\textwidth]{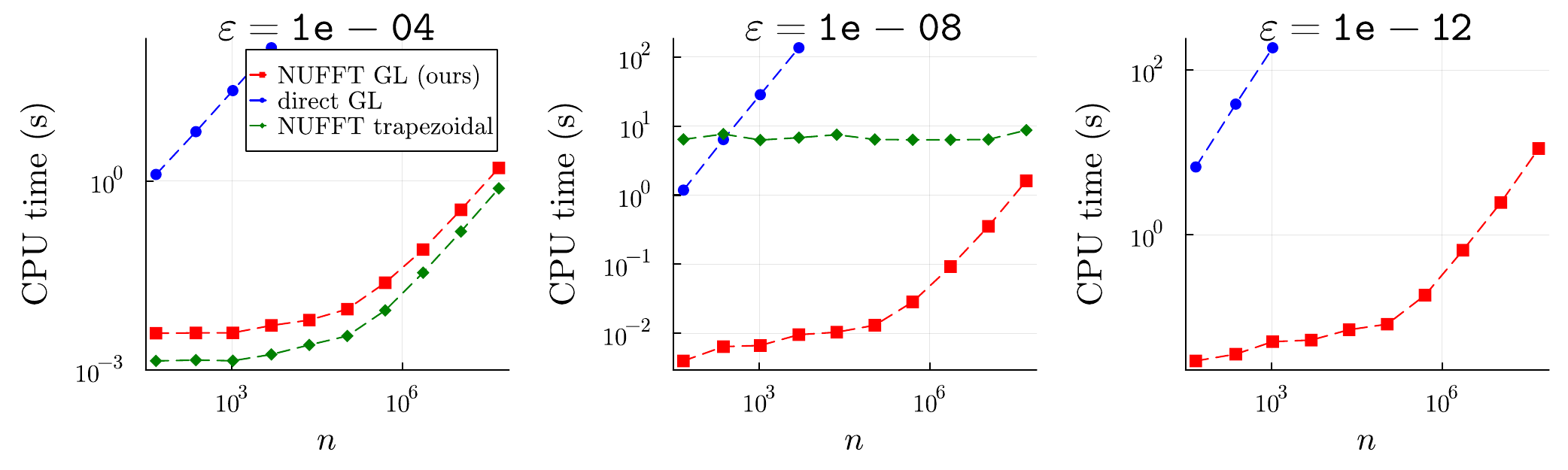}
    \caption{A comparison of the runtime cost of computing kernel values from a
    Mat\'ern model with $\nu = 0.55, \alpha=0.5,$ and $\phi$ chosen so that
    $K(0) = 1$. For three tolerances $\epsilon$ we plot Gauss-Legendre
    quadrature with direct summation (blue), global trapezoidal quadrature with
    the NUFFT (green), and Gauss-Legendre quadrature with the NUFFT (red). }
    \label{fig:timing}
\end{figure}
}{}

%% file: src/lidar.tex
In this section we demonstrate the practical value of our framework with an
application to high-frequency vertical wind profiles. The US Department of
Energy's Atmospheric Radiation Measurement (ARM) program offers a large
collection of freely available measurements collected at field sites across the
country, and in this work we look at the Doppler LiDAR-based vertical wind
profiles made at the main field site in the Southern Great Plains (SGP)
collection \citep{muradyan2020, newsom2012}. These measurements of vertical
profiles are made at a temporal resolution of approximately one second and a
spatial resolution of $30$m, thus providing particularly high resolution in
time. However, aside from some special small segments across the several years'
worth of data, the measurements have frequent interruptions due to horizontal
sweeps made by the sensor, occasional long pauses for various reasons, or other
momentary instrument-based delays. For this reason, unless one focuses on very
narrow time intervals of approximately $12-13$ minutes or is uninterested in
studying high-frequency structure of the process, 
approaches like ours for continuous-time models which are applicable to
irregularly sampled data become necessary.

\begin{remark}
  It is worth emphasizing that using a continuous-time process model for
  high-frequency data, even if that data is regularly gridded, can be valuable
  because it makes studying fine-scale properties such as mean-square
  differentiability more direct. Recall that if a continuous-time process $Z(t)$
  has spectral density $S(\omg)$, then the regularly measured ``time series"
  $Z_{\delta}(j) = Z(\delta j)$ has the aliased spectral density
  $S^{(\delta)}(\omg) = \delta^{-1} \sum_{\ell \in \Z} S\big(\delta^{-1}(\omg +
  \ell)\big)$ on $[-1/2,\, 1/2]$ (see, e.g., Section $3.6$ of
  \cite{stein1999interpolation}). While one could still study the
  differentiability of the process with a parametric model for the SDF
  $S^{(\delta)}$ and use some missing data method for time series models,
  modeling in the aliased spectral space and exchanging the Fourier integral we
  study here for a slowly converging, difficult to accelerate infinite series is
  not an appealing tradeoff.
\end{remark}

\subsection{Preliminary analysis}

In this work, we investigate the same six days that were studied in
\cite{geoga2021flexible}, but by lifting the limitation of gridded data, we can
examine a full hour of data on each of the days. Figure \ref{fig:lidardata03}
shows an example of one hour of measurements made at the altitude of $240$m.
\ifbool{figs}{
\begin{figure}[!ht]
  \centering
  \includegraphics[width=0.9\textwidth]{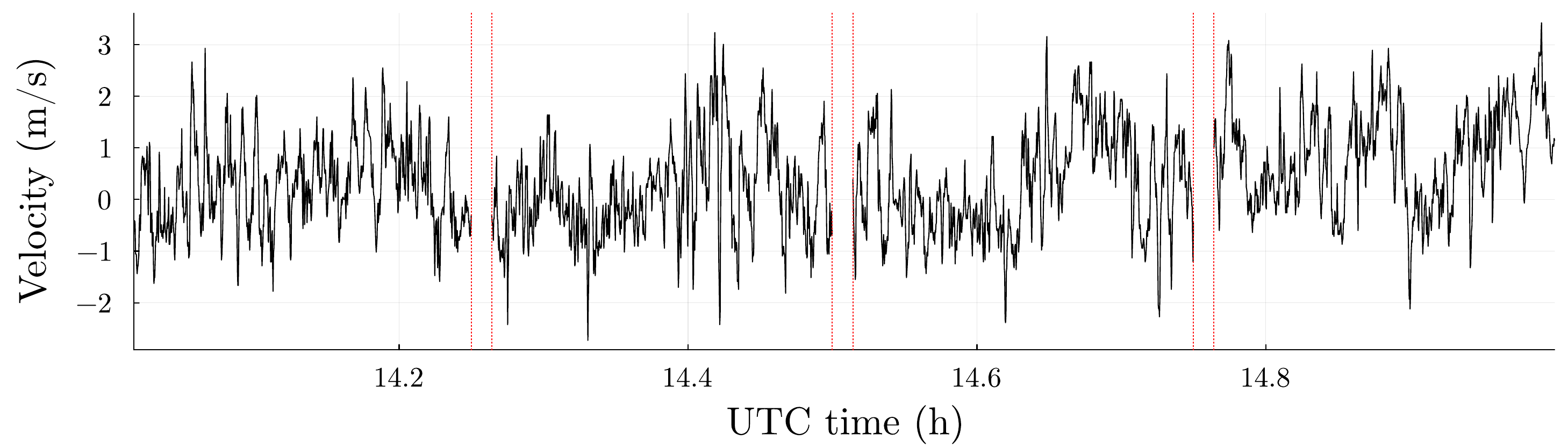}
  \caption{The Doppler LIDAR data from $1400$-$1500$ UTC on June $03$, $2015$,
  at an altitude of $240$m.  Vertical lines are given to emphasize the largest
  three gaps in measurements.}
  \label{fig:lidardata03}
\end{figure}
}{}
A practitioner looking to model this data may first compute a Whittle-type
estimator \citep{whittle1951} as an exploratory tool. Figure
\ref{fig:lidar_whittle} shows the result of Whittle estimation that is performed
by treating each of the six one-hour segments of data as i.i.d. samples,
breaking them into four segments based on the largest gaps and ignoring the
smaller irregularities, and computing one averaged periodogram per day. 
\ifbool{figs}{
\begin{figure}[!ht]
  \centering
  \includegraphics[width=\textwidth]{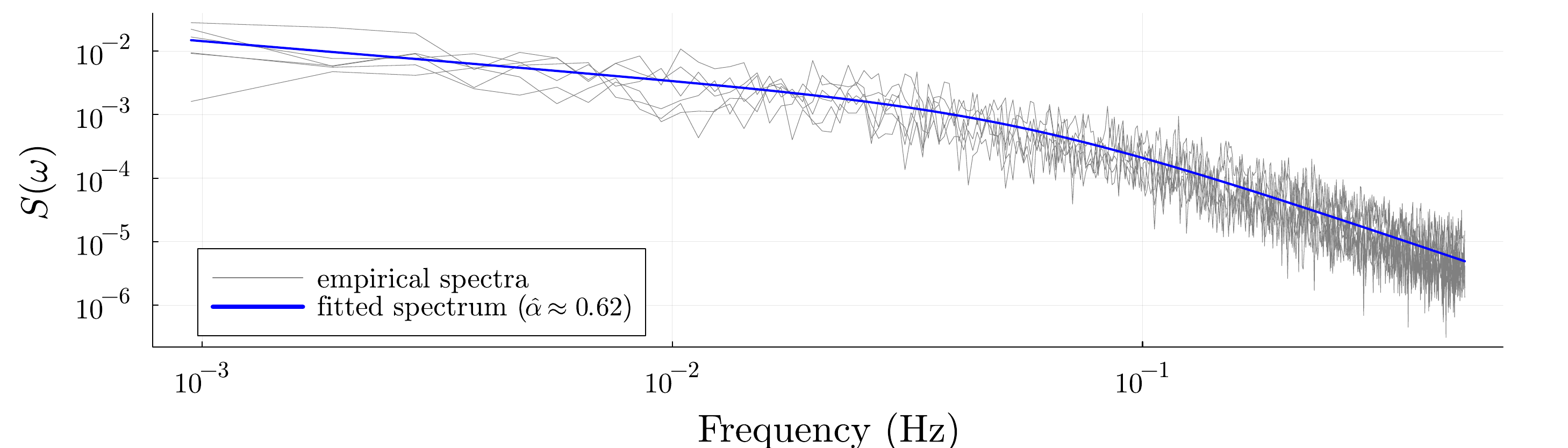}
  \caption{A Whittle MLE for the singular Mat\'ern model computed treating the
  six days of LiDAR data as i.i.d. samples and doing rudimentary gap and sample
  irregularity handling.}
  \label{fig:lidar_whittle}
\end{figure}
}{}
The Whittle MLE for this data implies a strong singularity of $\hat{\alpha}
\approx 0.62$, corresponding to a process with significant long-range dependence
even when segments are constrained to $12-13$ minutes, which limits the lowest
observable frequencies. 

\subsection{Parameter estimation methodology}

Motivated by this exploratory analysis, in this section we individually fit each
of the six days of data as a continuous-time process using the singular Mat\'ern
model. In all cases, the MLE indicates a strong singularity comparable to the
one obtained by the Whittle estimator. We additionally fit a standard Mat\'ern
model to compare log-likelihoods, and in all cases the log-likelihood of the
data is materially improved by adding the singularity. In the body of this
manuscript we focus on discussing one day of data, but the supplementary
material provides full diagnostics for all six days that were fitted.


All estimates discussed below and in the supplement were obtained using the
Artelys KNITRO optimizer \citep{byrd2006} with the sequential quadratic
programming (SQP) algorithm. In this spirit of Fisher scoring, we use the
\emph{expected} Fisher information matrix, given by
\begin{equation*} 
  \mathcal{I}(\bth)_{jk} = \frac{1}{2} \text{tr}
  \left(
    \bS_{\bth}^{-1} 
    \left[
    \frac{\partial}{\partial \theta_j}
    \bS_{\bth}
    \right]
    \bS_{\bth}^{-1} 
    \left[
    \frac{\partial}{\partial \theta_k}
    \bS_{\bth}
    \right]
  \right),
\end{equation*}
as a proxy for the Hessian of the log-likelihood. This matrix has the benefit of
being computable with only first derivatives of the covariance function, and has
meaningfully better performance than a general-purpose BFGS approximation
\citep{geoga2020scalable,
guinness2021gaussian,geoga2021flexible,beckman2023scalable}.

\subsection{Results}

We now discuss in detail the estimation results for the data on June $03$, one
of the six days that was studied. Analogous figures and tables for the other
five days can be found in the supplemental material. The results from fitting
singular and standard Mat\'ern models with a nugget are summarized by point
estimates and terminal likelihood values in Table
\ref{tab:lidar_matern_mles_03}. Along with point estimates, this table provides
standard deviations implied by the expected Fisher information matrix. These
implied uncertainties are often informative, but should be interpreted with care
because the regularity conditions under which the expected Fisher matrix
converges to the asymptotic precision of the MLE do not hold in many spatial
modeling settings \citep{stein1999interpolation}.
\ifbool{figs}{
\begin{table}[!ht]
  \renewcommand{\arraystretch}{1.3}
  \centering
  \begin{tabular}{|c|cc|}
  \hline
  \input{./figures/03_lidar_matern_mles.tex}
  \\ 
  \hline 
  \end{tabular}
  \caption{MLE estimates and terminal negative log-likelihoods for the standard
  Mat\'ern (fixed $\alpha=0$) and the singular Mat\'ern model for data on June
  $03$. When possible, expected Fisher matrix-implied standard deviations are
  provided in
  parentheses.}
  \label{tab:lidar_matern_mles_03}
\end{table}
}{}

The most obvious observation from Table \ref{tab:lidar_matern_mles_03} is that
the singular Mat\'ern model has a materially better log-likelihood than the
standard Mat\'ern model. A more scientifically interesting observation pertains
to the implied smoothness of the process: noting that
\begin{equation*} 
  \phi^2 |\omega|^{-\alpha} (\rho^2 + \omega^2)^{-\nu - \frac{1}{2}}
  \sim
  \omega^{-\alpha - 2\nu - 1}
\end{equation*}
at high frequencies, we see that the implied decay rate of the singular Mat\'ern
spectral density using the estimated $(\hat{\nu}, \hat{\alpha}) \approx (0.69,
0.78)$ is $\hat{\beta} = \hat{\alpha} + 2\hat{\nu} + 1 \approx 3.06$. Recall
that a process with spectral density $S$ is mean-square differentiable if and
only if $\int \omg^2 S(\omg) < \infty$, which in this notation is equivalent to
$\beta > 3$. Therefore, under the singular model where $\alpha$ is estimated,
the process is mean-square differentiable, whereas in the standard Mat\'ern
model the estimate of $\hat{\nu} = 0.79$ gives the decay $\hat{\beta} =
2\hat{\nu} + 1 \approx 2.58$ which implies that it is not. The differentiability
of these measurements below the atmospheric boundary layer (ABL) where
convective forces are dominant and the process exhibits chaotic behavior has
been a question of interest in several prior applications
\citep{geoga2021flexible, geoga2023scalable}, with parameter estimates often
being very borderline. The differentiability under the singular Mat\'ern model
is an indication that the long-range dependence parameter $\alpha$ may be
valuable in disentangling low-frequency and high-frequency behavior in such
processes, and agrees with the more recent work that uses continuous-time models
in kernel-space \citep{geoga2023scalable}.

A related observation that is particularly interesting from a theoretical
perspective is that, to the degree that the expected Fisher information matrices
can be trusted to serve as proxies for the precision of the MLE, the smoothness
and singularity parameters $\nu$ and $\alpha$ are jointly resolved reasonably
well. Table \ref{tab:lidar_matern_mles_03} shows that the implied uncertainties
for both of those parameters are small, whereas the parameters $\phi$ and $\rho$
which now most affect moderate frequencies are much less well-resolved. It is
well-known that in most fixed-domain asymptotic regimes in fewer than four
dimensions, no individual parameter in the Mat\'ern model can be estimated
consistently \citep{stein1999interpolation,zhang2004inconsistent}, and in light
of this observation the high uncertainty in $\phi$ and $\rho$ is not surprising.
For the singular Mat\'ern model in particular, however, $\rho$ serves the
important purpose of
making $S$ bounded at the origin, so that the singularity parameter $\alpha$ can
be disentangled from the effect of $\nu$ on the tail decay in the model
$|\omg|^{-\alpha} S(\omg)$.
\ifbool{figs}{ 
\begin{figure}[!ht]
  \centering
  \includegraphics[width=0.9\textwidth]{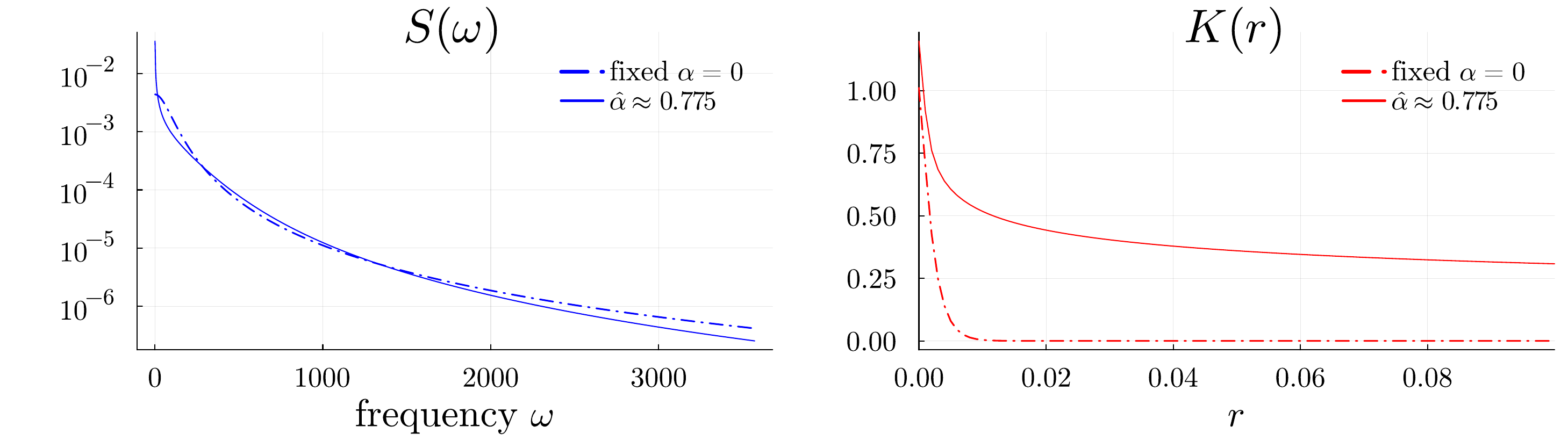}
  \caption{Left: Model-implied spectral densities for the data on June $03$ and
  the parameter estimates in Table \ref{tab:lidar_matern_mles_03}. Right: the
  corresponding model-implied kernels. }
  \label{fig:lidar_matern_sdfs_03}
\end{figure}
}{}

To understand what these estimates imply about the spectral densities and
covariances of the processes, Figure \ref{fig:lidar_matern_sdfs_03} shows the
MLE-implied spectra and kernel values from both models. 
As can be seen, the singular model moves much of the spectral mass into the
singularity near the origin, and the more rapid decay in the tails of the
singular Mat\'ern spectral density is clearly visible. The plot of the implied
kernel values in the center of Figure \ref{fig:lidar_matern_sdfs_03} shows a
more dramatic difference, as the estimated standard Mat\'ern kernel is highly
concentrated at the origin, while the singular Mat\'ern kernel displays slow
decay. 
Finally, Figure \ref{fig:lidar_matern_paths_03} shows sample paths from the two
MLE-implied models computed with the same white noise forcing. As Table
\ref{tab:lidar_matern_mles_03} and Figure \ref{fig:lidar_matern_sdfs_03} would
suggest, the additional high-frequency information in the standard Mat\'ern
model is quite prominent.
\ifbool{figs}{
\begin{figure}[!ht]
  \centering
  \includegraphics[width=0.9\textwidth]{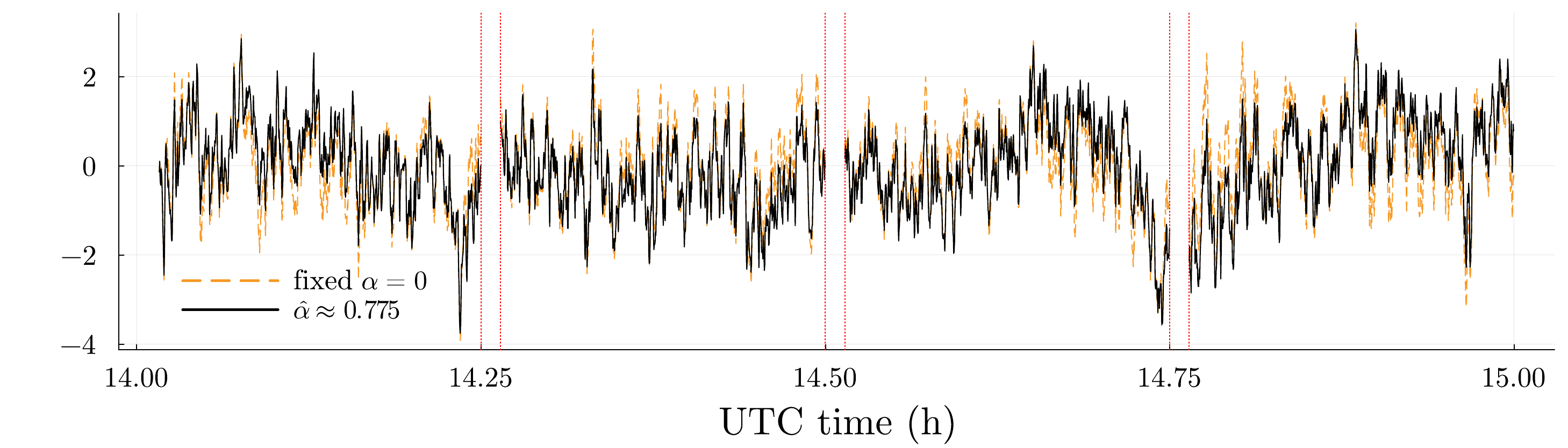}
  \caption{Sample paths with identical white noise forcing of the processes
  implied by the standard Mat\'ern MLE (orange dashes) and the singular Mat\'ern
  MLE (solid black), showing the noticeable affect of the more rapid spectral
  decay implied by the singular Mat\'ern model.}
  \label{fig:lidar_matern_paths_03}
\end{figure}
}{}

%% file: figures/03_lidar_matern_mles.tex
&fixed $ \alpha=0 $&fitted $\alpha$\\
\hline
$ \ell(\hat{\bth}) $&-1898.42&-1948.35\\
\hline
$ \phi $&25.72 (4.182)&205.3 (95.429)\\
$ \rho $&100.0 (7.854)&351.2 (45.705)\\
$ \nu $&0.7947 (0.023)&0.6908 (0.058)\\
$ \alpha $&---&0.775 (0.072)

%% file: src/discussion.tex
In this work we introduce a numerical method for efficiently and accurately
evaluating the Fourier transform of spectral densities, even those that are
barely integrable due to singularities at the origin or slowly decaying tails.
Making this strategy computational practical requires several technical
observations. The first and most crucial is that high-order quadrature rules can
significantly accelerate convergence. The use of the NUFFT is critical for this
purpose, as it reduces the computational cost to quasilinear complexity in both
the number of quadrature nodes $m$ and the number of inter-observation distances
$n$. The second vital observation is that the covariance function $K$ is
generally resolved to a given tolerance $\epsilon$ more quickly at distances $r$
that are well-separated from the origin. Therefore one can adaptively increase
the width of panels being integrated as one reduces the highest frequency
oscillations in the integrands remaining to be resolved. This Nyquist-based
observation is essential to overcoming the difficulties of slowly decaying
spectral densities, which may take orders of magnitude longer to converge if one
were to use a more uniform quadrature rule. Finally, the design of an efficient
but precise stopping criterion based on both truncation error \emph{and} panel
contribution is imperative to keeping the routine performant but accurate for a
wide variety of spectral densities. Combining all of these observations and the
additional technical discussions provided in Section \ref{sec:method}, one can
evaluate covariance functions specified by these spectral densities at tens of
millions of locations in seconds on a laptop.

The computational cost of using this method to compute kernel values will
naturally be significantly higher than direct kernel evaluation when a simple
closed form exists, for example the standard Mat\'ern model. Yet for the vast
majority of spectral densities whose Fourier transforms are completely
unavailable in closed form, our method is the first and only option for fitting
such models to irregularly sampled data.
The code and details of the integration strategy can surely be improved, further
reducing the overall cost of this approach. But even now it makes many modeling
choices available that were previously effectively impossible. While the LiDAR
application in this work is focused on the singular Mat\'ern model, we again
remind the reader that the objective of this framework is to empower
practitioners to work with \emph{any} continuous, integrable spectral density.
The extremely slow decay of a long-memory covariance function is of course not
always desirable, and there are many other new ways to add valuable degrees of
freedom to parametric families of spectral densities. It is our hope that this
method and the accompanying software will inspire others to explore and discover
new parametric families of models useful to their own contexts.

We emphasize that our framework can be used in conjunction with a number of
existing methods for large scale and nonstationary GPs. First, because it
requires only a spectral density and a list of distances $r_j$ at which to
evaluate the corresponding covariance function, our method can be used with any
approximation strategy for which these distances are specified in advance.
Examples include Vecchia approximation
methods~\citep{vecchia1988estimation,stein2004approximating,
katzfuss2021general} and entry-based rank-structured covariance matrix
approximations~\citep{chen2023linear,geoga2020scalable,beckman2023scalable}.
Second, we note that while our approach is of course intrinsically limited to
evaluating stationary covariance functions, it is compatible with a number of
methods which construct nonstationary models from stationary ones, including
warping \citep{sampson1992nonparametric} and convolutional frameworks
\citep{paciorek2006spatial}.
\ifbool{figs}{
\begin{figure}
    \hspace{0.02\textwidth} 
    $\log_{10} S(\bm{\omega})$
    \hspace{0.24\textwidth}
    $K(\bm{r})$
    \hspace{0.26\textwidth}
    $Z(\bm{x})$ \\
    \centering
    \vspace{-\baselineskip}
    \includegraphics[width=\textwidth]{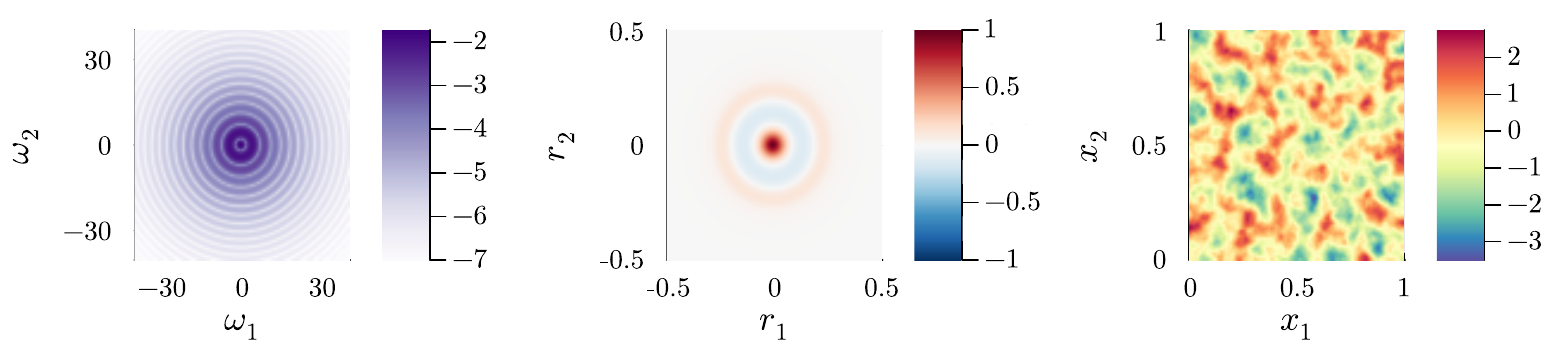}
    \caption{Log of spectral density (left), corresponding covariance function
    (center), and sample from the process (right) using the isotropic
    ``oscillatory'' Mat\'ern model (\ref{eq:oscillatory-matern}) in two
    dimensions. }
    \label{fig:example-2D}
\end{figure}
}{}

Finally, it is natural to ask how this methodology can be extended to multiple
dimensions. For any isotropic spectral density $S(\bw) = S(\norm{\bw})$ in
$\R^d$ one can integrate out the radial variables, resulting in the covariance
function
\begin{equation} \label{eq:higher-dimensions}
    K(r) = 2\pi \int_0^\infty S(\omega) J_{\frac{d}{2}-1}(2\pi r\omega) \, \omega^{\frac{d}{2}-1} \, d\omega
\end{equation}
where $J_\nu$ is the Bessel function of the first kind of order $\nu$. We can
compute this integral using the one-dimensional adaptive integration framework
just described, with the only modification being that the NUFFT is replaced by a
nonuniform fast Hankel transform. While there exist a number of fast Hankel
transform algorithms~\citep{cree1993algorithms, townsend2015fast}, we are not
aware of a fully nonuniform method in which $r$ and $\omega$ can be chosen
freely to accommodate the use of Gaussian quadrature rules for irregularly
sampled data. Without this fast transform, we are restricted to direct $\bO(nm)$
summation to compute each panel integral. Figure \ref{fig:example-2D} shows an
example of a 2D isotropic kernel computed with direct summation. The case of
anisotropic spectral densities is more complicated, and requires
multi-dimensional quadrature, but the modeling options it allows would be
enormous.

%% file: src/proofs_supp.tex
\begin{proof}[Proof of Lemma~\ref{lem:inc-gamma}]
    Using the integral representation 8.6.4 from~\cite{olver2010nist}, 
    \begin{align}
        \abs{\Gamma(-s, iy)}
        &= \abs{\frac{(iy)^{-s} \expp{-iy}}{\Gamma(s+1)} \int_0^\infty \frac{t^s \expp{-t}}{t+iy} \, dt} \\
        &\leq \frac{y^{-s}}{\Gamma(s+1)} \int_0^\infty \frac{t^s \expp{-t}}{\abs{t+iy}} \, dt \\
        &\leq \frac{y^{-s}}{\Gamma(s+1)} \int_0^\infty \frac{t^s \expp{-t}}{y} \, dt \\
        &= \frac{y^{-s-1}}{\Gamma(s+1)} \int_0^\infty t^s \expp{-t} \, dt \\
        &= y^{-s-1},
    \end{align}
    where $\Gamma(z) := \Gamma(0, z)$ is the usual gamma function. By a similar
    argument, we obtain
    \begin{align}
        \abs{\Gamma(-s, iy)}
        &\leq \frac{y^{-s}}{\Gamma(s+1)} \int_0^\infty \frac{t^s \expp{-t}}{t} \, dt \\
        &= \frac{y^{-s}}{\Gamma(s+1)} \Gamma(s) \\
        &= \frac{y^{-s}}{s}
    \end{align}
    using the fundamental property of the Gamma function $\Gamma(z+1) =
    z\Gamma(z)$.
\end{proof}

\begin{proof}[Proof of Theorem~\ref{thm:decay}]
    Define the covariance function $K := \Fr^{-1}\{S\}$ corresponding to $S$. As
    $S$ is integrable, $K$ is well defined and bounded
    \begin{equation}
        \abs{K(r)} 
        \leq \int_{-\infty}^\infty S(\omega) \abs{\cos(2\pi\omega r)} \dif \omega
        = \norm[{L^1(\R)}]{S}
        < \infty.
    \end{equation} 
    As $S \in C^2(\R)$ and $S', S'' \in L^1(\R)$, integrating by parts twice and
    applying the Riemann-Lebesgue lemma gives the standard decay rate $K(r) =
    o(r^{-2})$ for $r \to \infty$. Therefore $K$ is integrable with
    \begin{equation}
        M := \int_{-\infty}^\infty K(t) dt < \infty.
    \end{equation}
    Using the fact that
    \begin{equation}
        \Fr^{-1}\big\{\abs{\cdot}^{-\alpha}\big\}(r) 
        = (2\pi)^\alpha \frac{\Gamma(1-\alpha)}{\pi} \sin\left(\frac{\pi\alpha}{2}\right) |r|^{-1+\alpha}
    \end{equation}
    in the distributional sense and applying the convolution theorem, we obtain 
    \begin{align}
        K_\alpha(r) 
        &= \Fr^{-1} \Big\{\abs{\cdot}^{-\alpha} S(\cdot)\Big\}(r) \\
        &= \Big(\Fr^{-1} \big\{\abs{\cdot}^{-\alpha}\big\} * \Fr^{-1}\big\{S\big\}\Big)(r) \\
        &\propto \Big(\abs{\cdot}^{-1+\alpha} * K\Big)(r) \\
        &= \int_{-\infty}^\infty \abs{t}^{-1+\alpha} K(r-t) \dif t. \label{eq:conv-integral}
    \end{align}
    Take $0 < c < (1-\alpha)^{\frac{1}{2-\alpha}} < 1$. First consider the upper
    tail of the integral (\ref{eq:conv-integral}). Taylor expanding around $t=r$
    gives
    \begin{align}
        \int_{cr}^{\infty} t^{-1+\alpha} K(r-t) \dif t 
        &= \int_{cr}^{\infty} \Big( r^{-1+\alpha} - (1-\alpha)\xi^{-2+\alpha}(t - r) \Big) K(r-t) \dif t
    \end{align}
    for some $\xi$ between $r$ and $t$. This results in two terms. As $K$ is
    integrable, the first term gives the desired asymptotic behavior
    \begin{align} \label{eq:upper-conv-integral}
        r^{-1+\alpha} \int_{cr}^{\infty} K(r-t) \dif t
        &= r^{-1+\alpha} \Big( M - \int_{(1-c)r}^\infty K(u) \dif u \Big)
        \sim r^{-1+\alpha}.
    \end{align}
    As $\xi \geq cr$ and $K(r) = o(r^{-2})$ for $r \to \infty$, the second term
    decays with at least this asymptotic rate
    \begin{align}
        (1-\alpha) \int_{cr}^{\infty} \xi^{-2+\alpha}(t - r) K(r-t) \dif t
        \lesssim r^{-1+\alpha}
    \end{align}
    and by our choice of $c$, is strictly smaller in magnitude than
    (\ref{eq:upper-conv-integral}), which avoids cancellation. Next consider the
    lower tail of (\ref{eq:conv-integral})
    \begin{align}
        \int_{-\infty}^{-cr} (-t)^{-1+\alpha} K(r-t) \dif t 
        &\leq r^{-1+\alpha} \int_{(1+c)r}^{\infty} K(u) \dif u \label{eq:lower-conv-integral}
        \lesssim r^{-2+\alpha}
    \end{align}
    again due to the fact that $K(r) = o(r^{-2})$ for $r \to \infty$. Finally
    consider the central term in (\ref{eq:conv-integral}). Define the interval
    $I_r := [(1-c)r, (1+c)r]$. Then we have
    \begin{align} 
        \int_{-cr}^{cr} t^{-1+\alpha} K(r-t) \dif t
        &\leq \norm[{L^\infty(I_r)}]{K} \frac{2}{\alpha} (cr)^\alpha 
        \lesssim r^{-2+\alpha}
    \end{align} 
    as $\norm[{L^\infty(I_r)}]{K} = o(r^{-2})$ for $r \to \infty$.
    \end{proof}

%% file: src/numerics_supp.tex
\subsection{Error estimation for Mat\'ern model}

\renewcommand{\arraystretch}{1.1}

\begin{figure}[H]
  \centering
  \includegraphics[width=0.9\textwidth]{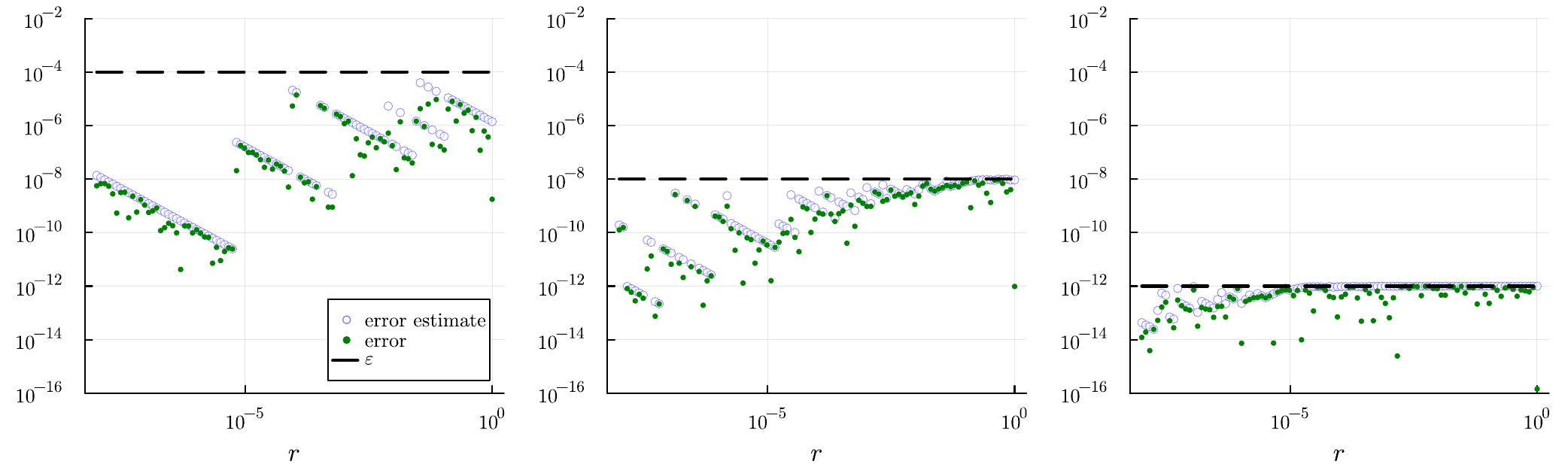}
  \caption{Pointwise errors and error estimates for various tolerances
  $\epsilon$ using a Mat\'ern model with $\nu = 0.51, \rho = 1,$ and $\phi$
  chosen so that $K(0) = 1$ with $m = 256$ quadrature nodes per panel,
  demonstrating the sharpness of the error control methods discussed in the
  previous section.} \label{fig:pointwise-error-nonsingular} 
\end{figure}

As a first test to validate the error estimation strategies described in
Section~\ref{sec:error-estimation}, we compare it to analytical evaluation of
the Mat\'ern covariance $K$ in the challenging regime $\nu = 0.51$, where the
spectral density $S$ decays slowly. We choose a moderate range parameter
$\rho=1$ and chose $\phi$ so that $K(0)=1$. For various tolerances $\epsilon$,
we compute 100 values of $K(r)$ for $r \in [\texttt{1e-8}, 1]$, along with the
corresponding error estimate for each $r$, given by the sum of the quadrature
and truncation error estimates discussed in Section~\ref{sec:error-estimation}.
See Figure~\ref{fig:pointwise-error-nonsingular}. Note that for all tested
tolerances $\epsilon$, all kernel values are computed to within $\epsilon$ of
the true value. In addition, the error estimates are generally accurate
indicators of the true errors, or at least act as upper bounds, which is to be
expected from the use of the truncation error bound
(\ref{eq:truncation-estimate}).

\newpage

\subsection{Comparison of methods for singular Mat\'ern model}

The following tables give precise values for several quantities of interest in
the numerical investigation of using the generalized hypergeometric function
representation of the singular Mat\'ern covariance function. For several values
of $\rho$ (including two that are not present in Figure
\ref{fig:singular_matern}), we provide the runtime cost of the kernel
evaluations, the absolute value of the kernel at $r = 1/2$ and $r=1$, and the
minimum eigenvalue of the covariance matrix assembled with those values. 

\begin{table}[H]
  \centering
  \begin{tabular}{|c|c|c|c|c|}
    \hline
    \input{./figures/singular_matern_compare_f64.tex}
    \\
    \hline
  \end{tabular}
  \caption{Full summary of numerical experiments using double precision.}
\end{table}

\begin{table}[H]
  \centering
  \begin{tabular}{|c|c|c|c|c|}
    \hline
    \input{./figures/singular_matern_compare_arb.tex}
    \\
    \hline
  \end{tabular}
  \caption{Full summary of numerical experiments using extended precision.}
\end{table}

\begin{table}[H]
  \centering
  \begin{tabular}{|c|c|c|c|c|}
    \hline
    \input{./figures/singular_matern_compare_qua.tex}
    \\
    \hline
  \end{tabular}
  \caption{Full summary of numerical experiments using our method.}
\end{table}

The standard double precision routines, while faster than our method, give
obviously incorrect results. Considering that in this study the kernel was
normalized so that $K(0) = 1$, one does not need to inspect the minimum
eigenvalue column very seriously when $K(1)$ is reported as being on the order
of $10^{14}$. The most relevant additional information that this table provides
is the demonstration that our method \emph{does} yield positive-definite
matrices in all cases.

%% file: figures/singular_matern_compare_f64.tex
&\multicolumn{4}{c|}{double precision} \\ 
\hline 
& time (s) & $|K(1/2)|$ & $|K(1)|$ & $\lambda_{\text{min}}(\bm{\Sigma}) $ \\ 
\hline 
$\rho=2$ & \num{4.726e-04} & \num{1.948e-01} & \num{1.017e-01} & \num{-1.208e-08} \\ 
$\rho=4$ & \num{4.350e-04} & \num{1.017e-01} & \num{5.922e-02} & \num{-1.556e-02} \\ 
$\rho=6$ & \num{5.078e-04} & \num{7.546e-02} & \num{2.373e+03} & \num{-1.980e+04} \\ 
$\rho=8$ & \num{5.707e-04} & \num{5.866e-02} & \num{1.507e+09} & \num{-6.930e+09} \\ 
$\rho=10$ & \num{6.412e-04} & \num{1.350e+00} & \num{6.758e+14} & \num{-2.516e+15} 

%% file: figures/singular_matern_compare_arb.tex
&\multicolumn{4}{c|}{extended precision} \\ 
\hline 
& time (s) & $|K(1/2)|$ & $|K(1)|$ & $\lambda_{\text{min}}(\bm{\Sigma}) $ \\ 
\hline 
$\rho=2$ & \num{1.825e+01} & \num{1.948e-01} & \num{1.017e-01} & \num{3.230e-11} \\ 
$\rho=4$ & \num{1.860e+01} & \num{1.017e-01} & \num{6.148e-02} & \num{7.312e-10} \\ 
$\rho=6$ & \num{1.885e+01} & \num{7.547e-02} & \num{2.180e+01} & \num{-2.497e+02} \\ 
$\rho=8$ & \num{1.904e+01} & \num{6.148e-02} & \num{1.006e+07} & \num{-1.018e+08} \\ 
$\rho=10$ & \num{1.923e+01} & \num{8.240e-02} & \num{4.165e+12} & \num{-3.438e+13} 

%% file: figures/singular_matern_compare_qua.tex
&\multicolumn{4}{c|}{our method} \\ 
\hline 
& time (s) & $|K(1/2)|$ & $|K(1)|$ & $\lambda_{\text{min}}(\bm{\Sigma}) $ \\ 
\hline 
$\rho=2$ & \num{9.260e-01} & \num{1.948e-01} & \num{1.017e-01} & \num{2.931e-11} \\ 
$\rho=4$ & \num{9.825e-01} & \num{1.017e-01} & \num{6.145e-02} & \num{7.305e-10} \\ 
$\rho=6$ & \num{9.128e-01} & \num{7.547e-02} & \num{4.613e-02} & \num{4.533e-09} \\ 
$\rho=8$ & \num{1.118e+00} & \num{6.145e-02} & \num{3.768e-02} & \num{1.654e-08} \\ 
$\rho=10$ & \num{9.042e-01} & \num{5.246e-02} & \num{3.222e-02} & \num{4.513e-08} 

%% file: src/lidar_supp.tex
\subsection{Model fits for LiDAR data: June 2} \label{sec:lidar_02}

\renewcommand{\arraystretch}{1.3}

\begin{figure}[!ht]
  \centering
  \includegraphics[width=0.7\textwidth]{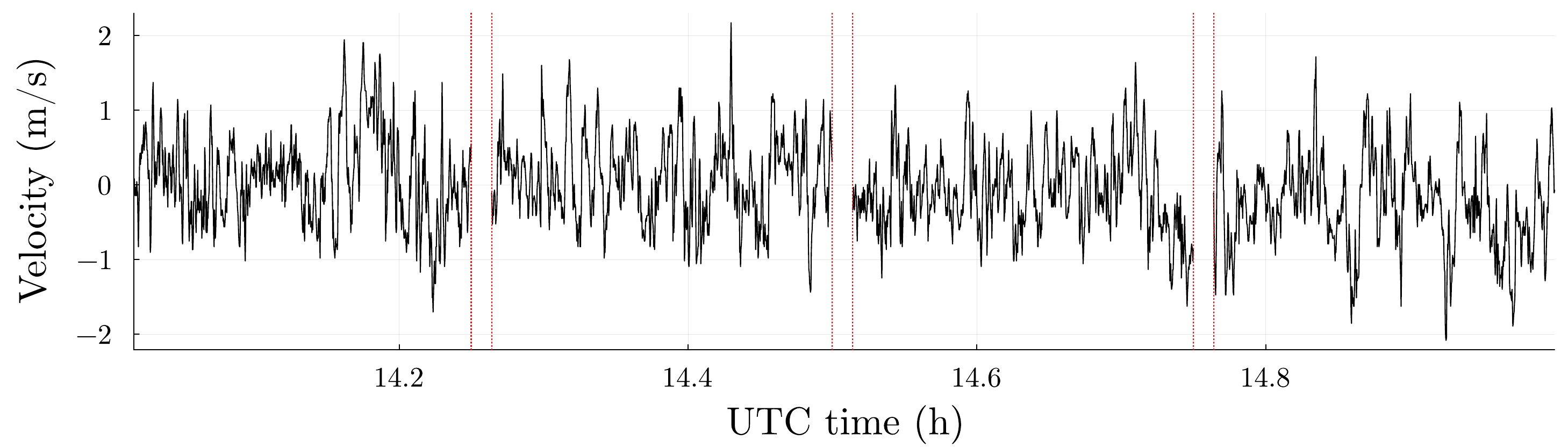}
  \caption{The Doppler LiDAR data from 1400-1500 UTC on June $02$, $2015$, with
  vertical lines to emphasize the largest three gaps.}
\end{figure}

\begin{figure}[!ht]
  \centering
  \includegraphics[width=0.9\textwidth]{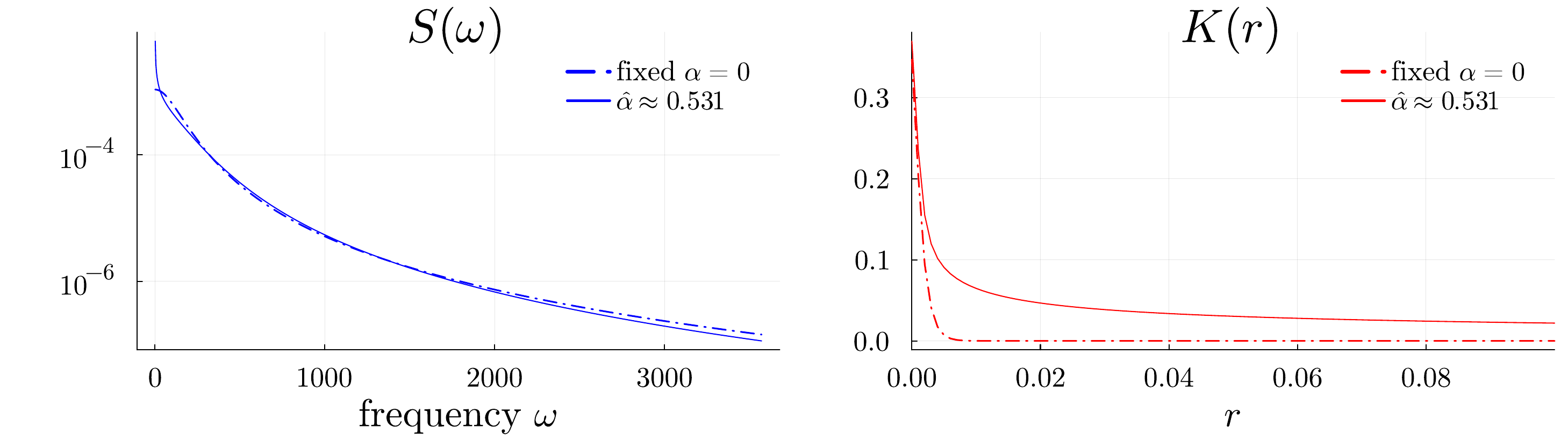}
  \caption{Left: model-implied SDFs for the data on June $02$ and the parameter
  estimates shown below. Center: the corresponding model-implied kernels. Right:
  sample paths from each of the model-implied kernel using the same white noise
  forcing.}
\end{figure}

\begin{table}[!ht]
  \footnotesize
  \centering
  \begin{tabular}{|c|cc|}
    \hline
    \input{./figures/02_lidar_matern_mles.tex}
    \\
    \hline
  \end{tabular}
  \caption{A summary of the terminal log-likelihood of the data in
  each modeling case as well as point estimates with expected Fisher
  matrix-implied standard deviations for the data of June $02$.}
\end{table}

\newpage

\subsection{Model fits for LiDAR data: June 6} \label{sec:lidar_06}

\begin{figure}[!ht]
  \centering
  \includegraphics[width=0.8\textwidth]{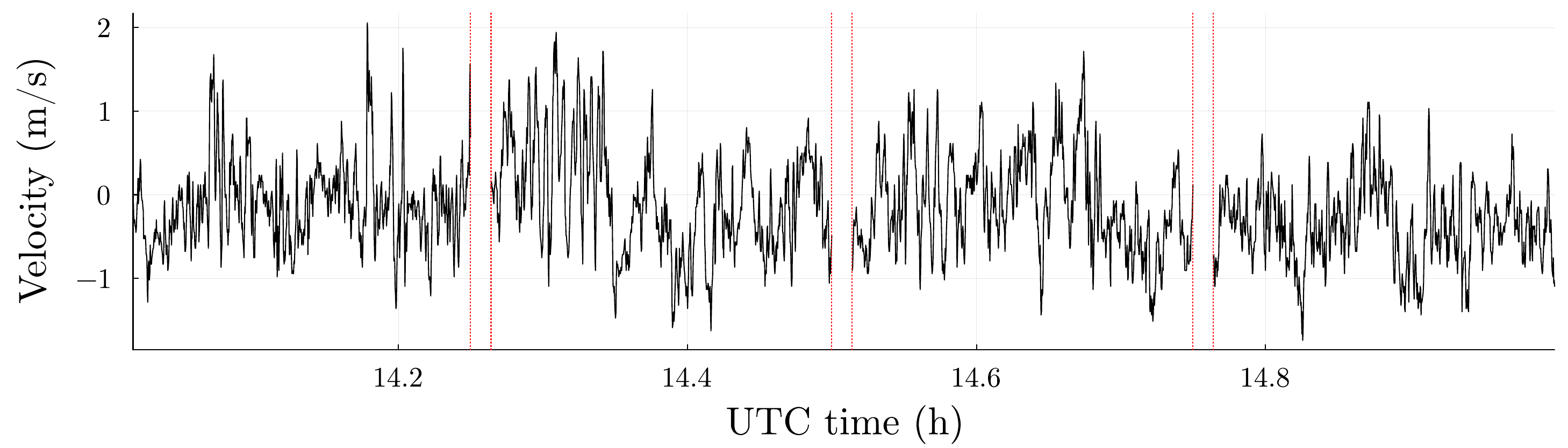}
  \caption{The Doppler LiDAR data from 1400-1500 UTC on June $06$, $2015$, with
  vertical lines to emphasize the largest three gaps.}
\end{figure}

\begin{figure}[!ht]
  \centering
  \includegraphics[width=\textwidth]{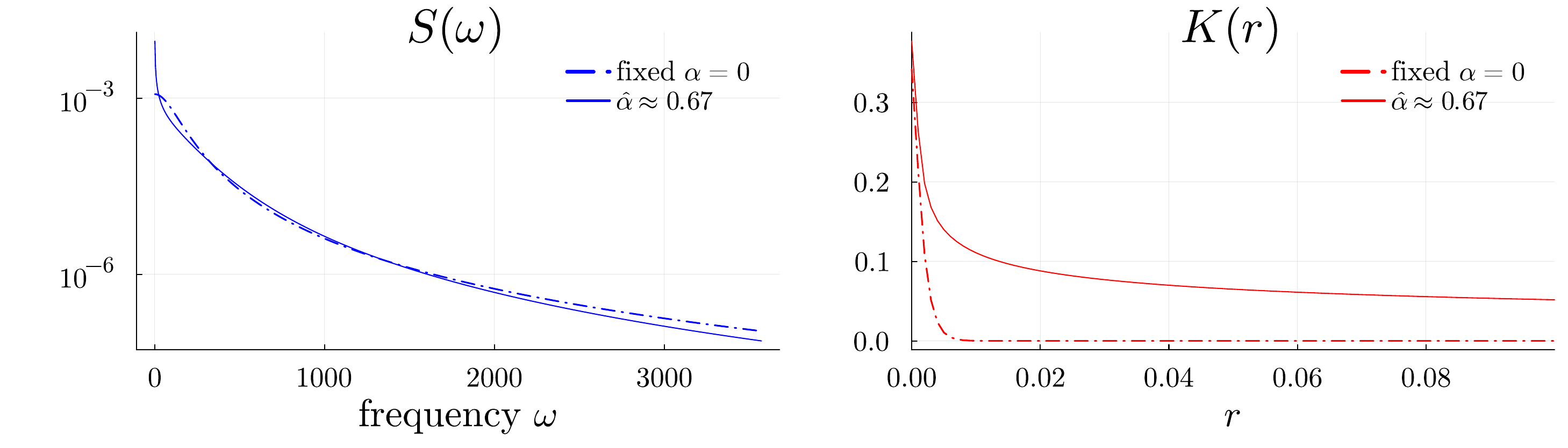}
  \caption{Left: model-implied SDFs for the data on June $06$ and the parameter
  estimates shown below. Center: the corresponding model-implied kernels. Right:
  sample paths from each of the model-implied kernel using the same white noise
  forcing.}
\end{figure}

\begin{table}[!ht]
  \footnotesize
  \centering
  \begin{tabular}{|c|cc|}
    \hline
    \input{./figures/06_lidar_matern_mles.tex}
    \\
    \hline
  \end{tabular}
  \caption{A summary of the terminal log-likelihood of the data in
  each modeling case as well as point estimates with expected Fisher
  matrix-implied standard deviations for the data of June $06$.}
\end{table}

\newpage

\subsection{Model fits for LiDAR data: June 20} \label{sec:lidar_20}

\begin{figure}[!ht]
  \centering
  \includegraphics[width=0.8\textwidth]{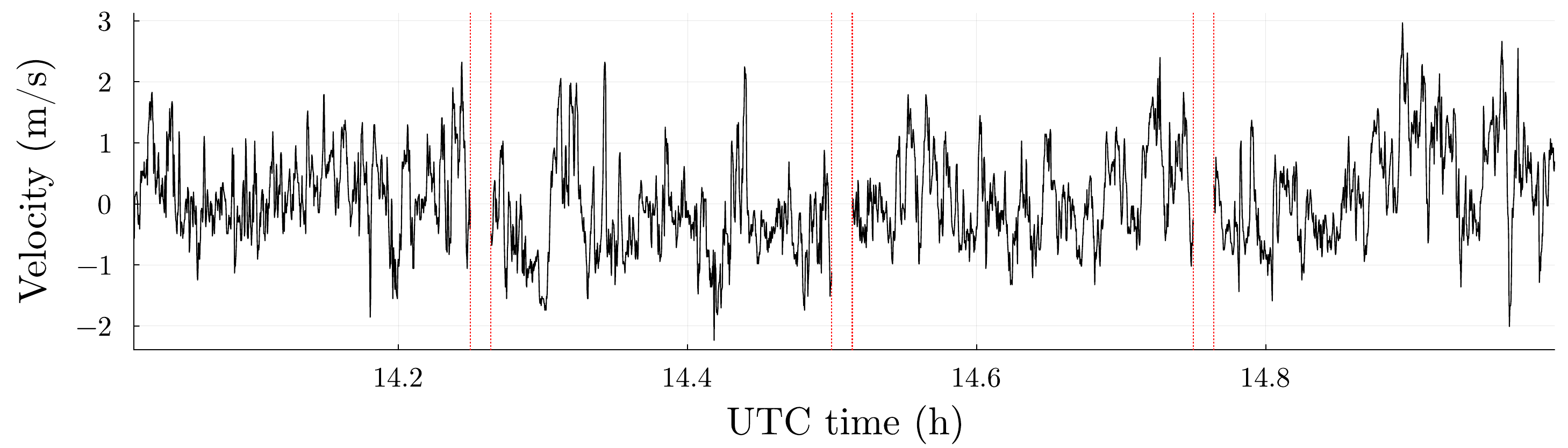}
  \caption{The Doppler LiDAR data from 1400-1500 UTC on June $20$, $2015$, with
  vertical lines to emphasize the largest three gaps.}
\end{figure}

\begin{figure}[!ht]
  \centering
  \includegraphics[width=\textwidth]{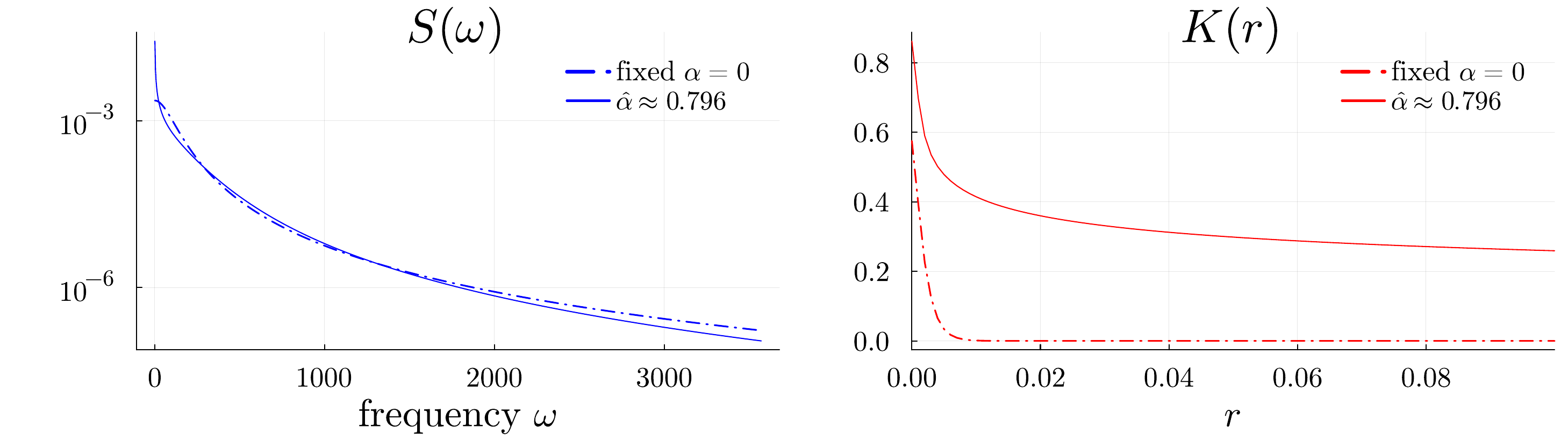}
  \caption{Left: model-implied SDFs for the data on June $20$ and the parameter
  estimates shown below. Center: the corresponding model-implied kernels. Right:
  sample paths from each of the model-implied kernel using the same white noise
  forcing.}
\end{figure}

\begin{table}[!ht]
  \footnotesize
  \centering
  \begin{tabular}{|c|cc|}
    \hline
    \input{./figures/20_lidar_matern_mles.tex}
    \\
    \hline
  \end{tabular}
  \caption{A summary of the terminal log-likelihood of the data in
  each modeling case as well as point estimates with expected Fisher
  matrix-implied standard deviations for the data of June $20$.}
\end{table}

\newpage

\subsection{Model fits for LiDAR data: June 24} \label{sec:lidar_24}

\begin{figure}[!ht]
  \centering
  \includegraphics[width=0.8\textwidth]{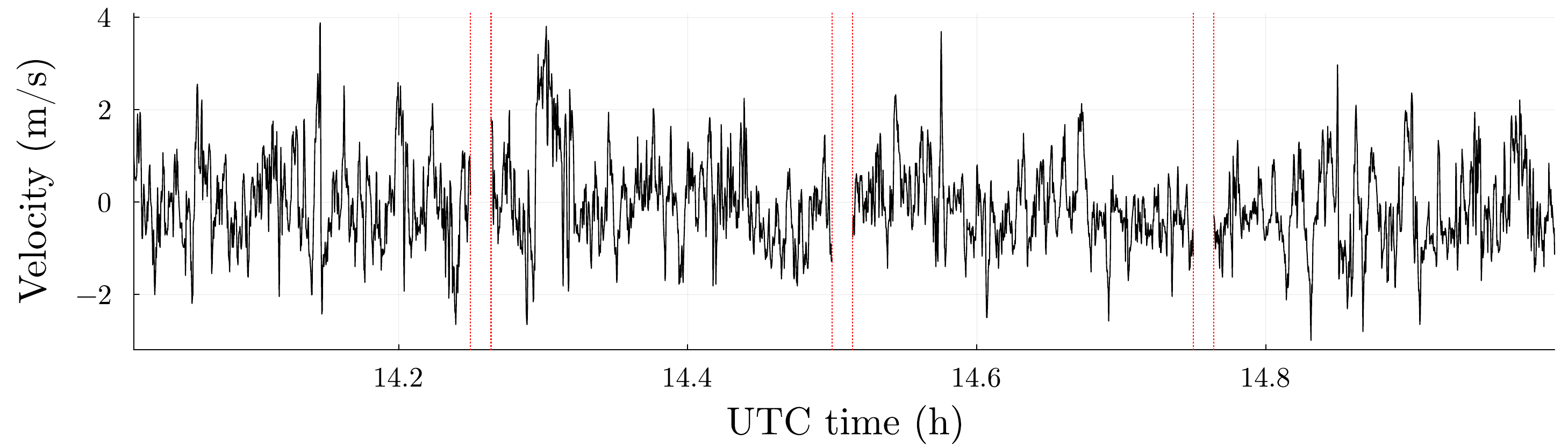}
  \caption{The Doppler LiDAR data from 1400-1500 UTC on June $24$, $2015$, with
  vertical lines to emphasize the largest three gaps.}
\end{figure}

\begin{figure}[!ht]
  \centering
  \includegraphics[width=\textwidth]{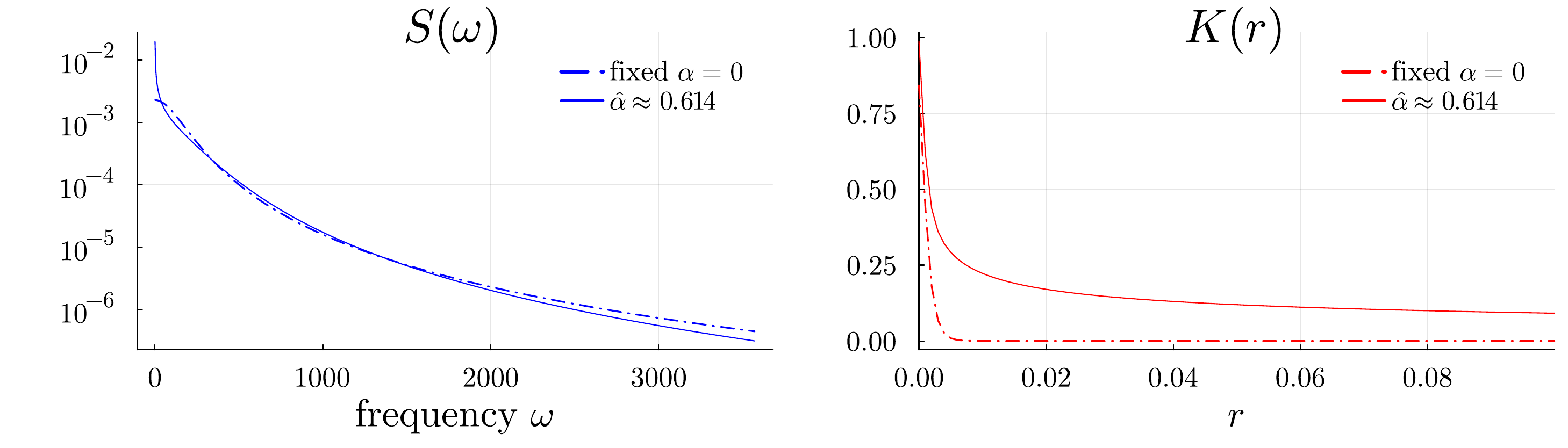}
  \caption{Left: model-implied SDFs for the data on June $24$ and the parameter
  estimates shown below. Center: the corresponding model-implied kernels. Right:
  sample paths from each of the model-implied kernel using the same white noise
  forcing.}
\end{figure}

\begin{table}[!ht]
  \footnotesize
  \centering
  \begin{tabular}{|c|cc|}
    \hline
    \input{./figures/24_lidar_matern_mles.tex}
    \\
    \hline
  \end{tabular}
  \caption{A summary of the terminal log-likelihood of the data in
  each modeling case as well as point estimates with expected Fisher
  matrix-implied standard deviations for the data of June $24$.}
\end{table}

\newpage

\subsection{Model fits for LiDAR data: June 28} \label{sec:lidar_28}

\begin{figure}[!ht]
  \centering
  \includegraphics[width=0.8\textwidth]{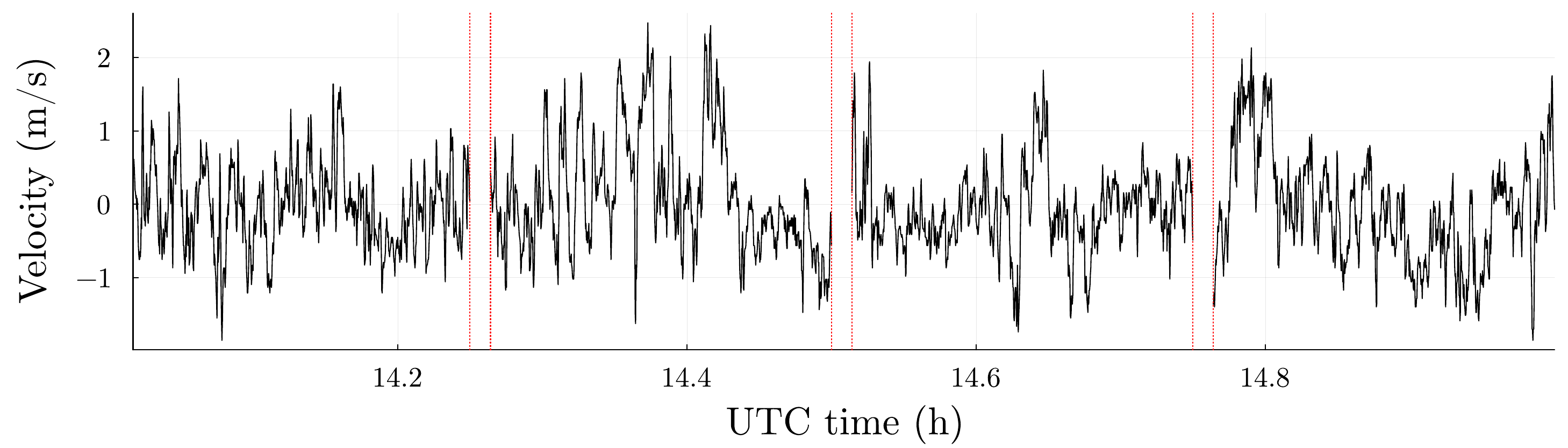}
  \caption{The Doppler LiDAR data from 1400-1500 UTC on June $28$, $2015$, with
  vertical lines to emphasize the largest three gaps.}
\end{figure}

\begin{figure}[!ht]
  \centering
  \includegraphics[width=\textwidth]{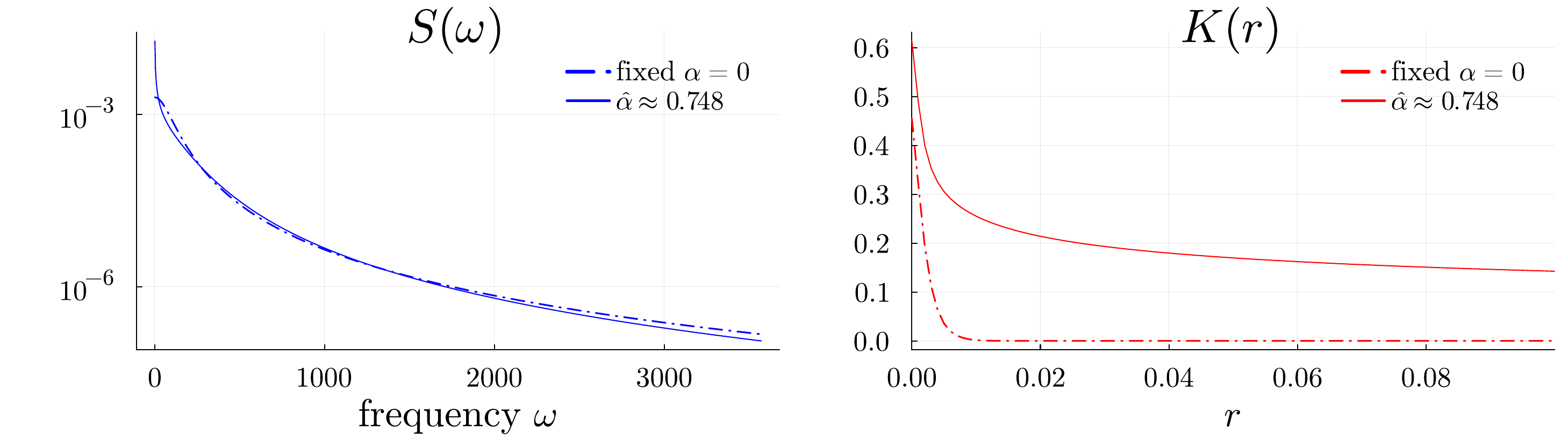}
  \caption{Left: model-implied SDFs for the data on June $28$ and the parameter
  estimates shown below. Center: the corresponding model-implied kernels. Right:
  sample paths from each of the model-implied kernel using the same white noise
  forcing.}
\end{figure}

\begin{table}[!ht]
  \footnotesize
  \centering
  \begin{tabular}{|c|cc|}
    \hline
    \input{./figures/28_lidar_matern_mles.tex}
    \\
    \hline
  \end{tabular}
  \caption{A summary of the terminal log-likelihood of the data in
  each modeling case as well as point estimates with expected Fisher
  matrix-implied standard deviations for the data of June $28$.}
\end{table}

%% file: figures/02_lidar_matern_mles.tex
&fixed $ \alpha=0 $&fitted $\alpha$\\
\hline
$ \ell(\hat{\bth}) $&-3267.49&-3283.12\\
\hline
$ \phi $&40.35 (8.393)&109.2 (39.915)\\
$ \rho $&153.1 (10.595)&277.6 (32.599)\\
$ \nu $&0.9146 (0.029)&0.7844 (0.05)\\
$ \alpha $&---&0.5314 (0.084)

%% file: figures/06_lidar_matern_mles.tex
&fixed $ \alpha=0 $&fitted $\alpha$\\
\hline
$ \ell(\hat{\bth}) $&-3600.7&-3638.64\\
\hline
$ \phi $&41.37 (8.193)&218.1 (97.376)\\
$ \rho $&140.5 (9.624)&335.3 (39.221)\\
$ \nu $&0.9356 (0.028)&0.8269 (0.056)\\
$ \alpha $&---&0.6696 (0.076)

%% file: figures/20_lidar_matern_mles.tex
&fixed $ \alpha=0 $&fitted $\alpha$\\
\hline
$ \ell(\hat{\bth}) $&-2975.98&-3003.8\\
\hline
$ \phi $&34.7 (6.144)&202.4 (86.377)\\
$ \rho $&115.1 (8.314)&323.2 (39.589)\\
$ \nu $&0.8865 (0.025)&0.7297 (0.055)\\
$ \alpha $&---&0.7965 (0.073)

%% file: figures/24_lidar_matern_mles.tex
&fixed $ \alpha=0 $&fitted $\alpha$\\
\hline
$ \ell(\hat{\bth}) $&-1467.18&-1481.26\\
\hline
$ \phi $&73.93 (16.889)&367.8 (182.718)\\
$ \rho $&176.7 (12.177)&368.0 (44.475)\\
$ \nu $&0.92 (0.032)&0.8301 (0.061)\\
$ \alpha $&---&0.614 (0.075)

%% file: figures/28_lidar_matern_mles.tex
&fixed $ \alpha=0 $&fitted $\alpha$\\
\hline
$ \ell(\hat{\bth}) $&-3399.46&-3420.24\\
\hline
$ \phi $&21.83 (3.617)&64.81 (21.275)\\
$ \rho $&102.7 (7.782)&251.3 (32.702)\\
$ \nu $&0.8374 (0.024)&0.6127 (0.048)\\
$ \alpha $&---&0.7477 (0.082)